\documentclass[twocolumn]{aastex63}
\def\actaa{Acta Astronomica}

\usepackage{amsmath}
\hypersetup{linkcolor=red,citecolor=blue,filecolor=green,urlcolor=magenta}

\newcommand{\feh}{$\mathrm{[Fe/H]}$}

\begin{document}

\shorttitle{Photometric Metallicity for RR Lyrae}
\shortauthors{Ngeow}

\title{Evaluating the $V$-band Photometric Metallicity with Fundamental Mode RR Lyrae in the Kepler Field}

\correspondingauthor{C.-C. Ngeow}
\email{cngeow@astro.ncu.edu.tw}

\author[0000-0001-8771-7554]{Chow-Choong Ngeow}
\affil{Graduate Institute of Astronomy, National Central University, 300 Jhongda Road, 32001 Jhongli, Taiwan}

\begin{abstract}
  The aim of this work is to evaluate the performance of photometric metallicity \feh, determined based on $V$-band light-curves photometrically transformed from the $gr$-band light-curves. We tested this by using a set of homogeneous sample of fundamental mode RR Lyrae located in the {\it Kepler} field. It was found that the color-term is necessary in such photometric transformation. We demonstrated that including the color-term the determined photometric \feh~are in good agreement to the spectroscopic \feh, either based on the calibrated or the transformed $V$-band light-curves. We also tested the impact of Blazhko RR Lyrae in determining the photometric \feh, and found that Blazhko RR Lyrae can give consistent photometric \feh. Finally, we derived independent $gVr$-band \feh-$\phi_{31}$-$P$ relations (where $\phi_{31}$ and $P$ are Fourier parameter and pulsation period, respectively) using our light-curves. The $V$-band relation is in good agreement with the most recent determination given in the literature.

\end{abstract}

%\keywords{ {\bf TBD}}

\section{Introduction}\label{sec1}

One of the important observed quantity, besides the pulsation periods $P$, for RR Lyrae is metallicity, commonly denoted as \feh. This is because the $V$-band absolute magnitude ($M_V$) for RR Lyrae is correlated with \feh. Also, \feh~is one of the independent parameters in the period-luminosity-metallicity (PLZ) relations (especially in the infra-red filters). For exemplary reviews on $M_V$-\feh~relation or PLZ relations, see \citet{sandage2006}, \citet{beaton2018}, \citet{bhardwaj2020}, and reference therein.

The best way to measure \feh~is via spectroscopic observations. Errors on the spectroscopically measured \feh~fall in the range of $\sim0.1$ to $\sim 0.3$~dex \citep[e.g., see][]{nemec2013}, in some cases these errors can even reach to $\sim 0.01$~dex level \citep[e.g., SX For in][]{crestani2021}. However, spectroscopic observations for RR Lyrae can be expensive or time-consuming. Alternatively, \feh~can be estimated based on the stellar systems (for examples, in globular clusters or dwarf galaxies) or environment (for example, in Galactic halo) at which the targeted RR Lyrae are belong to. Another way to estimate \feh~is using light-curve properties of RR Lyrae, such as amplitudes \citep[for examples, see][]{alcock2000,sandage2004,fabrizio2021} or Fourier parameter $\phi_{31}$.

Since the seminal paper of \citet{jurcsik1996}, who derived the \feh-$\phi_{31}$-$P$ relation in the $V$-band for ab-type (fundamental mode) RR Lyrae (hereafter RRab), a number of publications has derived a similar relation (some with additional parameters in such relation) in other filters, as well as for the c-type (first overtone) RR Lyrae. These works include: \citet{sandage2004} and \citet{morgan2007} in the $V$-band; \citet{smolec2005} and \citet{dekany2021} in the $I$-band; \citet{watkins2009}, \citet{sesar2010}, and \citet{oluseyi2012} in the Sloan Digital Sky Survey (SDSS) $g$- and/or $r$-band; \citet{nemec2011} and \citet{nemec2013} in the {\it Kepler} $K_p$-band; \citet{ngeow2016} in the $R_{PTF}$-band; \citet{iorio2021} in the {\it Gaia} $G$-band; \citet{mullen2021} in the {\it WISE} $W1$- and $W2$-band; and \citet{wu2006} for unfiltered or white-light observations. The RMS (root-mean-square) errors from these empirical relations vary from $\sim0.1$~dex \citep{nemec2013} to $\sim0.5$~dex \citep[in the {\it WISE} band,][]{mullen2021}.

Recently, the $V$-band \feh-$\phi_{31}$-$P$ relation was updated from two works. \citet{marvaz2016} updated the \citet{jurcsik1996} relation by using 7 globular clusters and 8 field RR Lyrae with high-resolution spectroscopic metallicity. Furthermore, \citet{mullen2021} re-derived the $V$-band \feh-$\phi_{31}$-$P$ relation based on a sample of $\sim 10^3$ field RRab with spectroscopic determined \feh. It is foreseen that the \citet{mullen2021} $V$-band relation will be widely applied in various studies on using ab-type RR Lyrae as distance tracers. On the other hand, the SDSS-like $(u)griz$ filters are becoming more popular in major synoptic sky surveys, including (but not limited to) the Pan-STARRS1 \citep{chambers2016}, the Zwicky Transient Facility \citep[ZTF,][]{bel19,gra19}, the SkyMapper Southern Survey \citep{onken2019}, the Dark Energy Survey \citep{des2016}, the HyperSuprime-Cam Subaru Strategic Program \citep{aihara2018}, and the Vera C. Rubin Observatory Legacy Survey of Space and Time \citep[LSST,][]{ivezic2019}. This implies that in order to apply the \citet{mullen2021} $V$-band relation, photometric transformations need to be applied to the $gr$-band data from these surveys to the $V$-band. In principle, such transformations could add extra uncertainties to the final estimated \feh.

Therefore, the goal of this work is to evaluate the performance and accuracy of such photometric transformations in the derivation of photometric \feh, using the $V$-band \feh-$\phi_{31}$-$P$ relation. Instead of relying on inhomogeneous data taken from literature, we intended to obtain homogeneous light-curve data using the same telescope and CCD camera on the same set of RR Lyrae, such that a differential comparison can be made. We selected 30 RRab stars located in the {\it Kepler} field \citep[taken from Table 7 in][]{nemec2013}, which possess homogeneous spectroscopic \feh~measured from high-resolution spectra. Section \ref{sec2} describes the time-series observations of these RR Lyrae. Photometry and photometric calibration of our light-curves data are presented in Section \ref{sec3}. These light-curves were then used to derive their corresponding Fourier parameter $\phi_{31}$ as mentioned in Section \ref{sec4}. Performance on the photometric \feh~from using the $V$-band light-curves and the transformed light-curves are tested in Section \ref{sec5}. We have also derived a set of \feh-$\phi_{31}$-$P$ relations based on our light-curves in Section \ref{sec6}, followed by discussions and conclusions in Section \ref{sec7}. We reminded that the transmission curve for $V$-band filter is lied in between the $g$- and the $r$-band filters, hence we only observed our targeted RR Lyrae in these filters.

\section{Observations and Image Reduction} \label{sec2}

Time-series observations of the 30 targeted RR Lyrae in the {\it Kepler} field were carried out using the 0.41-meter SLT telescope located at Lulin Observatory. This telescope is a $f/8.4$ Ritchey-Chr\'{e}tien telescope and it is equipped with a Andor iKon-L936 CCD camera, providing a pixel scale of $0.79\arcsec/$pixel. Queue observations were executed, via commercial software {\tt MaxIm DL} and {\tt ACP Observatory Control Software}, from 18 June 2019 to 24 November 2021 (weather permitted) in $gVr$ filters. Depends on the brightness of the targeted RR Lyrae, exposure time varies between 2 to 300~second in all filters. After removing problematic images (due to bad seeings or weather, tracking problems, etc), the number of $gVr$ sequence ranged from $\sim120$ to $\sim144$ for all of the 30 RR Lyrae. Subroutines in {\tt IRAF} (Image Reduction and Analysis Facility; version 2.16)\footnote{{\tt IRAF} is distributed by the National Optical Astronomy Observatories, which are operated by the Association of Universities for Research in Astronomy, Inc., under cooperative agreement with the National Science Foundation. See \url{https://ascl.net/9911.002}} were used to reduce these images, including bias and dark subtractions, as well as flat-fieldings. Astrometric calibration on the reduced images were done using the {\tt astrometry.net}\footnote{\url{https://astrometry.net/} or \url{https://ascl.net/1208.001}} \citep{lang2010} software suite.

\section{Photometric Calibration} \label{sec3}

For each of our targeted RR Lyrae, we constructed a reference catalog by merging the Pan-STARRS1 Data Release 1 (DR1) photometric data \citep{chambers2016,flewelling2020} and the $UBV$ photometric catalog published in \citet[][hereafter the $UBV$ catalog]{everett2012}. A search area with a size of $27\arcmin \times 27\arcmin$ centered at each targeted RR Lyrae was adopted to query the Pan-STARRS1 DR1 photometric data. We applied a number of selection criteria to select only the non-varying stellar sources in the merged reference catalogs. Further details of the adopted selection criteria were given in the Appendix \ref{appxa}. These merged reference catalogs were then cross-matched to the catalogs generated from the {\tt SExtractor}\footnote{\url{https://www.astromatic.net/software/sextractor/} or \url{https://ascl.net/1010.064}} \citep[version 2.25.0,][]{bertin1996} on all reduced images. The popular $MAG\_AUTO$ implemented in {\tt SExtractor} was adopted for measuring the instrumental magnitudes. Hence, each images we have a catalog containing both of the $gr$-band and $BV$-band photometry from the Pan-STARRS1 and the $UBV$ catalog, respectively, for the reference stars, as well as their instrumental magnitudes.

The photometric calibration was done using the following set of equations \citep[for example, see][]{masci2019}:

\begin{eqnarray}
  g^{PS1} - g^{\mathrm{instr}} & = &  ZP_g + C_g(g^{PS1}-r^{PS1}), \\
  r^{PS1} - r^{\mathrm{instr}} & = &  ZP_r + C_r(g^{PS1}-r^{PS1}), \\
  V - V^{\mathrm{instr}} & = &  ZP_V + C_V(B^{EHK}-V^{EHK}),
\end{eqnarray}

\noindent where $m^{PS1\ \mathrm{or}\ EHK}$ are magnitudes from published catalogs (either Pan-STARRS1 or $UBV$ catalog), and $m^{\mathrm{instr}}$ are instrumental magnitudes. An iterative $2\sigma$-clipping linear regression, implemented in {\tt astropy}, was used to fit these equations to determine the $ZP_m$ and $C_m$ coefficients.

Since our SLT observations did not include the $B$ filter, we employed the photometric transformations given in \citet{tonry2012} to calibrate the $(B-V)$ colors. We adopted the linear transformation between Johnson and Pan-STARRS1 photometric system from Table 6 of \citet{tonry2012}: $B - g^{PS1} = 0.213 + 0.587(g^{PS1}-r^{PS1})$ and $V - r^{PS1} = 0.006 + 0.474(g^{PS1}-r^{PS1})$. Then, the color transformation is found to be:

\begin{eqnarray}
  (B - V) & = & 0.207 + 1.113(g^{PS1}-r^{PS1}).
\end{eqnarray}

\noindent Finally, we can transform the calibrated $r^{PS1}$ magnitudes to the $V$-band magnitude via \citet{tonry2012} transformation. The transformed $V$-band magnitudes are denoted as $VT$:

\begin{eqnarray}
  VT & = & r^{PS1} + 0.006 + 0.474(g^{PS1}-r^{PS1}).
\end{eqnarray}

\begin{figure*}
  \epsscale{1.15}
  \plottwo{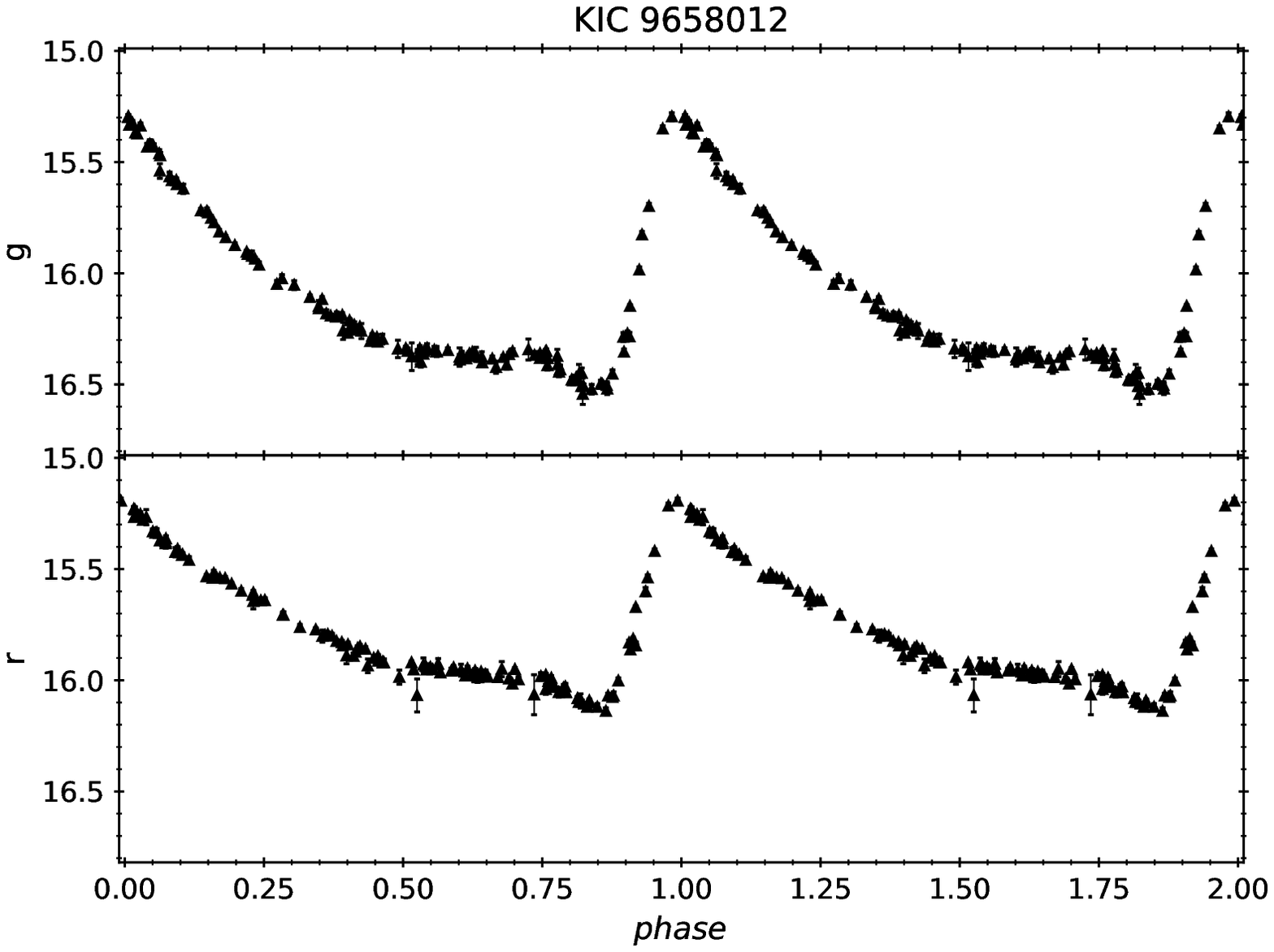}{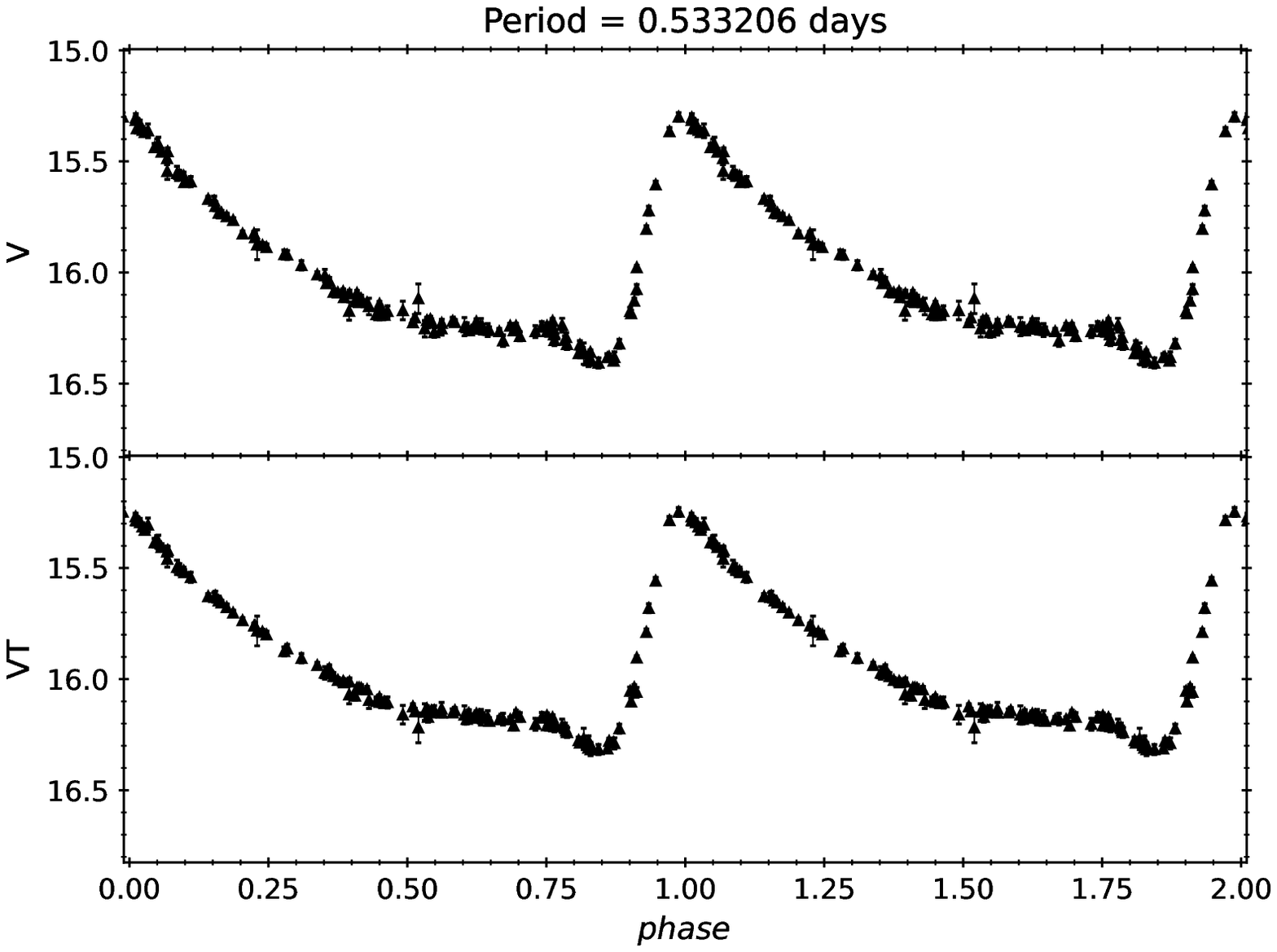}
  \caption{Calibrated $grV$-band light-curves and the transformed $VT$-band light-curve for a non-Blazhko RR Lyrae. The error bars include errors from the instrumental magnitudes and the calibration processes.}
  \label{fig_lc}
\end{figure*}

\begin{deluxetable*}{lccccccccc}
  %\movetableright=-1in
  \tabletypesize{\scriptsize}
  \tablecaption{$grV$-band Calibrated Light-Curves for 30 ab-Type RR Lyrae in the {\it Kepler} Field\label{tab_lc}}
  \tablewidth{0pt}
  \tablehead{
    \colhead{KIC} &
    \colhead{HJD$_g$} &
    \colhead{$g$} &
    \colhead{$\sigma_g$} &
    \colhead{HJD$_r$} &
    \colhead{$r$} &
    \colhead{$\sigma_r$} &
    \colhead{HJD$_V$} &
    \colhead{$V$} &
    \colhead{$\sigma_V$} 
  }
  \startdata
  3733346 & 2459094.1610997790 & 13.131 & 0.038 & 2459094.1621529846 & 12.807 & 0.019 & 2459094.1616437426 & 13.128 & 0.076 \\
  3733346 & 2458779.0534368511 & 13.140 & 0.024 & 2458779.0544553185 & 12.767 & 0.013 & 2458779.0539576588 & 13.009 & 0.050 \\
  3733346 & 2458775.0634470126 & 13.159 & 0.018 & 2458775.0644654804 & 12.799 & 0.011 & 2458775.0639678198 & 13.072 & 0.026 \\
  3733346 & 2459019.2116095843 & 13.122 & 0.016 & 2459019.2126744250 & 12.786 & 0.014 & 2459019.2121651536 & 13.022 & 0.033 \\
  $\cdots$  &  $\cdots$ & $\cdots$ &  $\cdots$  &  $\cdots$ & $\cdots$ &   $\cdots$  &  $\cdots$ & $\cdots$ &  $\cdots$ \\
  \enddata
  \tablecomments{Table \ref{tab_lc} is published in its entirety in the machine-readable format. A portion is shown here for guidance regarding its form and content. HJD is the helio-Julian date at the mid-points of exposures.}
\end{deluxetable*}

Applying equation (1) to (5) to calibrate the instrumental magnitudes requires the $(g^{PS1}-r^{PS1})$ colors of the targeted stars to be known. In case for our SLT observations with a sequence of $gVr$ observations, the separation between the $gVr$-band exposures within a sequence is always less than 30~minutes (with a median of 6.2~minutes), and hence we assume the photometry obtained from the near-simultaneous $gr$-band observations is equivalent to the $(g-r)$ color at the time of observations. Combining equation (1) \& (2), the instrumental colors can be calibrated to the Pan-STARRS1 photometric system via the following equation:

\begin{eqnarray}
  (g^{PS1}-r^{PS1}) & = & \frac{ZP_g-ZP_r + (g^{\mathrm{instr}} - r^{\mathrm{instr}})}{1-C_g+C_r}.
\end{eqnarray}

\noindent The calibrated $(g^{PS1}-r^{PS1})$ colors can be applied back to equation (1) to (5) to calibrate the $grV$ and $VT$-band photometry. An example of the calibrated $grV$-band and the transformed $VT$-band light-curve is shown in Figure \ref{fig_lc}. All of the calibrated $grV$-band light-curves are provided in Table \ref{tab_lc}. Photometric errors given in Table \ref{tab_lc} and shown in Figure \ref{fig_lc} include the errors from the instrumental magnitudes and the propagated errors from the calibration. Typical errors on $ZP_{g,r,V}$ are $\sim 0.006$~mag, $\sim 0.004$~mag, and $\sim 0.013$~mag, respectively. Similarly, the typical errors on $C_{g,r,V}$ are $\sim 0.012$~mag, $\sim 0.007$~mag, and $\sim 0.018$~mag, respectively.

\section{Fourier Parameters $\phi_{31}$} \label{sec4}

\begin{figure*}
  \gridline{\fig{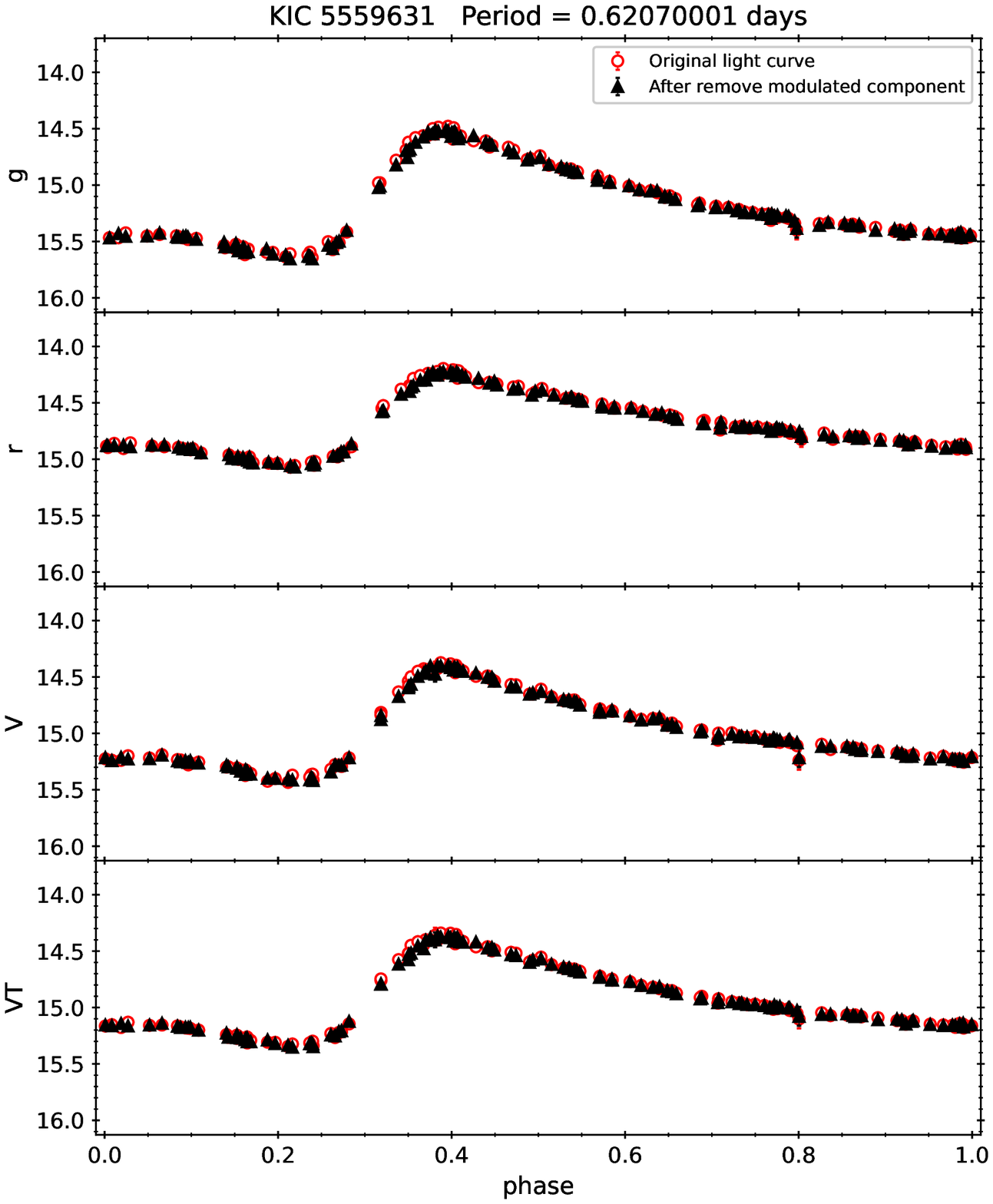}{0.32\textwidth}{(a) No signs of Blazhko modulation seen on the light-curves.}
    \fig{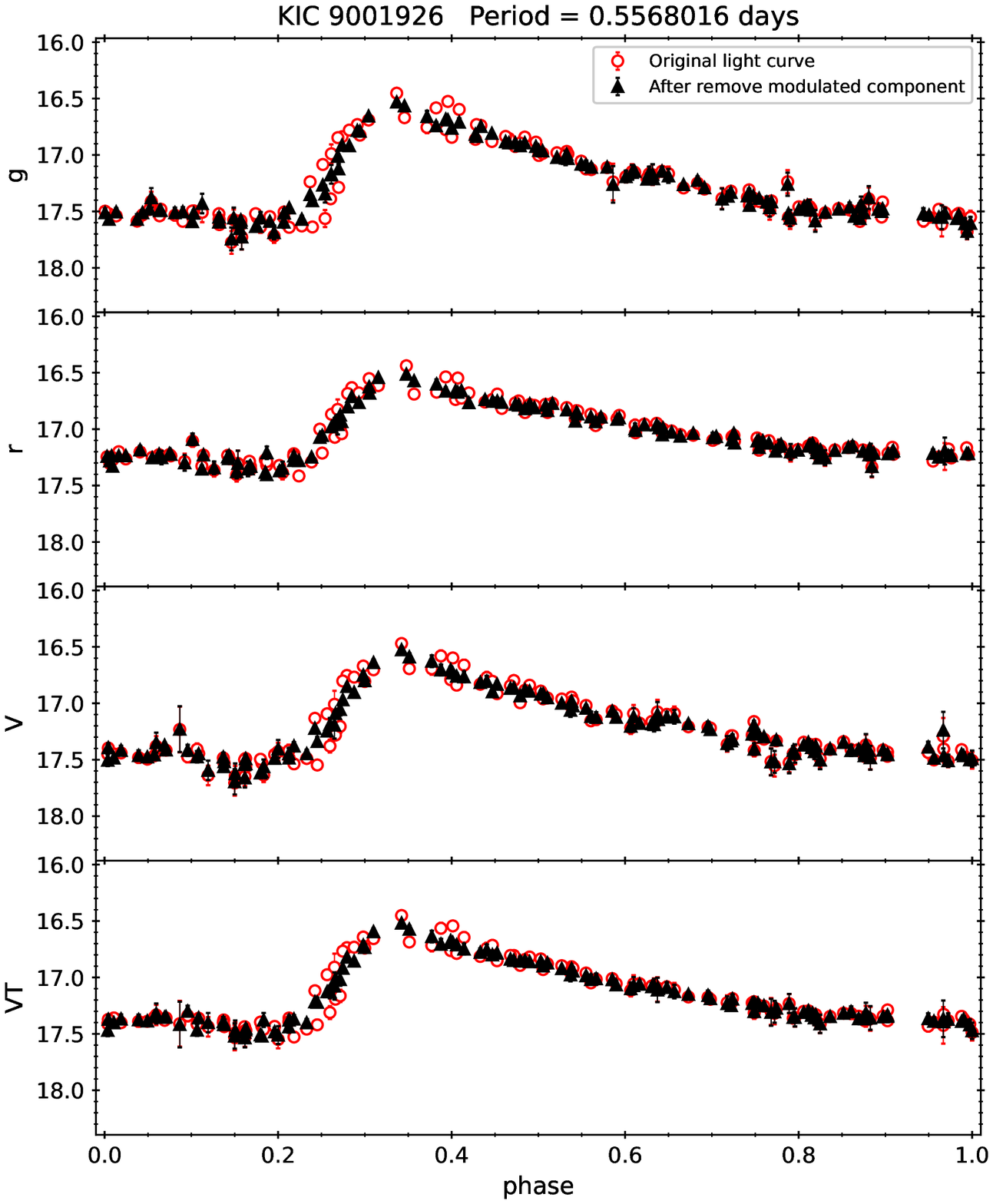}{0.32\textwidth}{(b) Improvements seen on the light-curves after removing the Blazhko modulation.}
    \fig{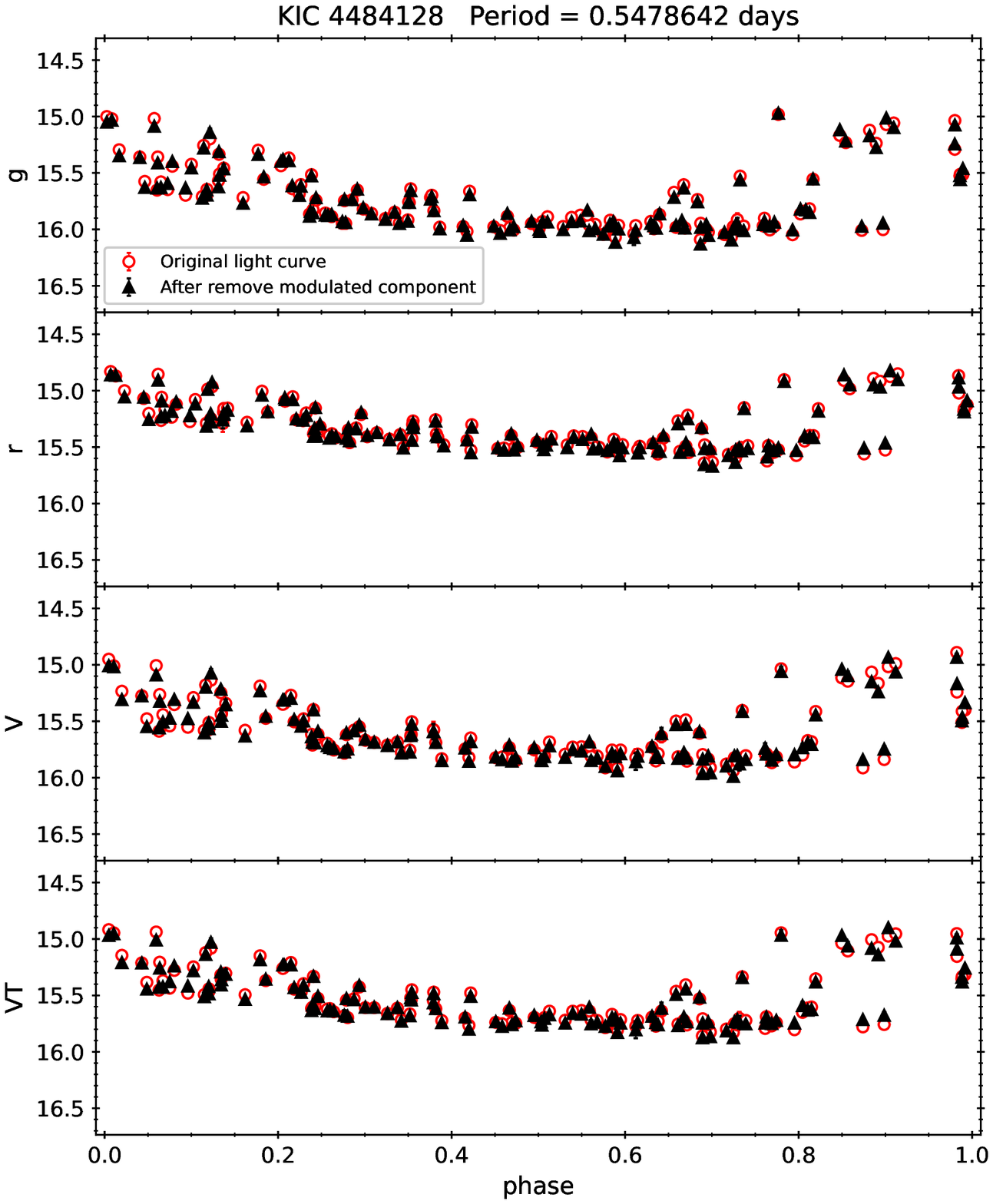}{0.32\textwidth}{(c) No improvements seen on the light-curves even after removing the Blazhko modulation.}
  }
  \caption{Example light-curves for Blazhko RR Lyrae before and after removing the Blazhko modulation. The error bars include errors from the instrumental magnitudes and the calibration processes. Panel (a) shows an example that no clear sign of Blazhko modulation seen in the light-curves. Panel (b) and (c) are examples where improvements and no improvements were seen on the light-curves after removing the Blazhko modulation, respectively. }
  \label{fig_blazhko}
\end{figure*}

The $V$-band photometric \feh~given in \citet{mullen2021} is:

\begin{eqnarray}
  \mathrm{[Fe/H]_V}  & = & -1.22[\pm0.01] - 7.60[\pm0.24] (P-0.58)  \nonumber \\
   &  & +1.42[\pm0.05](\phi_{31}^s - 5.25),
\end{eqnarray}

\noindent with an RMS of 0.41~dex. Precise $P$ for our targeted RR Lyrae are available from \citet{nemec2013} based on the {\it Kepler} observations. Hence, we only need to determine the Fourier parameter $\phi_{31}$ from our calibrated $V$-band light-curves to obtain the \feh$_V$. 

\begin{deluxetable*}{lllcccccl}
  %\movetableright=-1in
  \tabletypesize{\scriptsize}
  \tablecaption{Basic Information and the Derived $\phi^s_{31}$ for Our Targeted RR Lyrae\label{tab_fourier}}
  \tablewidth{0pt}
  \tablehead{
    \colhead{KIC} &
    \colhead{$P$\tablenotemark{a}} &
    \colhead{$P_{BL}$\tablenotemark{a}} &
    \colhead{$\phi^s_{31}$} &
    \colhead{$\phi^s_{31}$} &
    \colhead{$\phi^s_{31}$} &
    \colhead{$\phi^s_{31}$} &
    \colhead{\feh$_{\mathrm{spec}}$\tablenotemark{a}} &
    \colhead{Other Name} \\
    \colhead{} &
    \colhead{(days)} &
    \colhead{(days)} &
    \colhead{($g$-band)} &
    \colhead{($r$-band)} &
    \colhead{($V$-band)} &
    \colhead{($VT$-band)} &
    \colhead{(dex)} &
    \colhead{}
  }
  \startdata
  \multicolumn{9}{c}{Non-Blazhko RR Lyrae} \\
  3733346  & 0.6820264  & $\cdots$ & $4.857\pm0.028$ & $5.156\pm0.028$ & $4.941\pm0.051$ & $5.006\pm0.053$ & $-2.54\pm0.11$ & NR Lyr \\
  3866709  & 0.47070609 & $\cdots$ & $4.811\pm0.028$ & $5.107\pm0.029$ & $4.842\pm0.036$ & $4.950\pm0.037$ & $-1.13\pm0.09$ & V715 Cyg \\
  5299596  & 0.5236377  & $\cdots$ & $5.420\pm0.030$ & $5.840\pm0.031$ & $5.538\pm0.038$ & $5.606\pm0.039$ & $-0.42\pm0.10$ & V782 Cyg \\
  6070714  & 0.5340941  & $\cdots$ & $5.672\pm0.030$ & $6.090\pm0.031$ & $5.804\pm0.038$ & $5.853\pm0.039$ & $-0.05\pm0.10$ & V784 Cyg \\
  6100702  & 0.4881457  & $\cdots$ & $5.356\pm0.018$ & $5.706\pm0.024$ & $5.489\pm0.038$ & $5.514\pm0.039$ & $-0.16\pm0.09$ & $\cdots$ \\
  6763132  & 0.5877887  & $\cdots$ & $4.843\pm0.019$ & $5.108\pm0.019$ & $4.933\pm0.041$ & $4.969\pm0.042$ & $-1.89\pm0.10$ & NQ Lyr \\
  6936115  & 0.52739847 & $\cdots$ & $4.618\pm0.015$ & $4.878\pm0.016$ & $4.704\pm0.030$ & $4.750\pm0.031$ & $-1.98\pm0.09$ & FN Lyr \\
  7030715  & 0.68361247 & $\cdots$ & $5.308\pm0.017$ & $5.718\pm0.020$ & $5.452\pm0.033$ & $5.498\pm0.034$ & $-1.33\pm0.08$ & $\cdots$ \\
  7742534  & 0.4564851  & $\cdots$ & $4.627\pm0.020$ & $4.819\pm0.024$ & $4.639\pm0.028$ & $4.683\pm0.029$ & $-1.28\pm0.20$ & V368 Lyr \\
  9508655  & 0.5942369  & $\cdots$ & $4.894\pm0.019$ & $5.095\pm0.023$ & $5.015\pm0.033$ & $4.989\pm0.034$ & $-1.83\pm0.12$ & V350 Lyr \\
  9591503  & 0.5713866  & $\cdots$ & $4.857\pm0.016$ & $5.125\pm0.017$ & $4.942\pm0.029$ & $4.985\pm0.030$ & $-1.66\pm0.12$ & V894 Cyg \\
  9658012  & 0.533206   & $\cdots$ & $4.865\pm0.018$ & $5.217\pm0.022$ & $4.965\pm0.026$ & $5.032\pm0.027$ & $-1.28\pm0.14$ & $\cdots$ \\
  9717032  & 0.5569092  & $\cdots$ & $4.900\pm0.041$ & $5.099\pm0.048$ & $4.996\pm0.053$ & $5.020\pm0.055$ & $-1.27\pm0.15$ & $\cdots$ \\
  9947026  & 0.5485905  & $\cdots$ & $5.362\pm0.019$ & $5.862\pm0.024$ & $5.501\pm0.036$ & $5.576\pm0.038$ & $-0.59\pm0.13$ & V2470 Cyg \\
  10136240 & 0.5657781  & $\cdots$ & $4.863\pm0.023$ & $5.303\pm0.029$ & $4.975\pm0.036$ & $5.057\pm0.037$ & $-1.29\pm0.23$ & V1107 Cyg \\
  10136603 & 0.4337747  & $\cdots$ & $5.372\pm0.016$ & $5.676\pm0.022$ & $5.467\pm0.030$ & $5.507\pm0.031$ & $-0.05\pm0.14$ & V839 Cyg \\
  11802860 & 0.6872160  & $\cdots$ & $5.281\pm0.017$ & $5.519\pm0.022$ & $5.363\pm0.034$ & $5.396\pm0.035$ & $-1.33\pm0.09$ & AW Dra \\
  \multicolumn{9}{c}{Blazhko RR Lyrae} \\
  3864443  & 0.4869538  &  234.0   & $4.626\pm0.024$ & $4.571\pm0.031$ & $4.473\pm0.041$ & $4.577\pm0.042$ & $-1.66\pm0.13$ & V2178 Cyg \\
  4484128  & 0.5478642  &  92.14   & $4.888\pm0.032$ & $5.018\pm0.036$ & $4.861\pm0.041$ & $4.872\pm0.046$ & $-1.19\pm0.18$ & V808 Cyg \\
  5559631  & 0.62070001 &  27.667  & $5.251\pm0.016$ & $5.507\pm0.019$ & $5.335\pm0.024$ & $5.374\pm0.025$ & $-1.16\pm0.11$ & V783 Cyg \\
  6183128  & 0.561691   & 723.0    & $4.651\pm0.026$ & $4.933\pm0.031$ & $4.681\pm0.039$ & $4.756\pm0.040$ & $-1.44\pm0.16$ & V354 Lyr \\
  7198959  & 0.566788   &   39.20  & $4.786\pm0.093$ & $5.831\pm0.088$ & $5.273\pm0.173$ & $5.362\pm0.194$ & $-1.27\pm0.12$ & RR Lyrae \\
  7505345  & 0.4737027  &   31.05  & $4.584\pm0.025$ & $4.724\pm0.029$ & $4.656\pm0.040$ & $4.678\pm0.044$ & $-1.14\pm0.17$ & V355 Lyr \\
  7671081  & 0.5046123  &  123.7   & $4.860\pm0.033$ & $5.134\pm0.042$ & $4.771\pm0.047$ & $4.943\pm0.050$ & $-1.51\pm0.12$ & V450 Lyr \\
  9001926  & 0.5568016  &  71.6    & $4.917\pm0.040$ & $5.200\pm0.048$ & $4.945\pm0.058$ & $5.008\pm0.061$ & $-1.50\pm0.20$ & V353 Lyr \\
  9578833  & 0.5270283  &  62.84   & $5.003\pm0.025$ & $5.165\pm0.031$ & $5.084\pm0.038$ & $5.019\pm0.039$ & $-1.16\pm0.09$ & V366 Lyr \\
  9697825  & 0.5575765  &  52.07   & $4.884\pm0.026$ & $5.185\pm0.034$ & $4.971\pm0.041$ & $4.990\pm0.042$ & $-1.50\pm0.29$ & V360 Lyr \\
  10789273 & 0.48027971 &   54.0   & $4.702\pm0.015$ & $4.924\pm0.017$ & $4.795\pm0.025$ & $4.827\pm0.026$ & $-1.01\pm0.10$ & V838 Cyg \\
  11125706 & 0.6132200  &  40.23   & $5.478\pm0.050$ & $5.913\pm0.054$ & $5.707\pm0.106$ & $5.683\pm0.114$ & $-1.09\pm0.08$ & $\cdots$ \\
  12155928 & 0.43638507 &   51.999 & $4.618\pm0.012$ & $4.790\pm0.016$ & $4.701\pm0.027$ & $4.725\pm0.028$ & $-1.23\pm0.15$ & V1104 Cyg \\
  \enddata
  \tablenotetext{a}{Values adopted from \citet{nemec2013}.}
  \tablecomments{For Blazhko RR Lyrae, the $\phi^s_{31}$ values were derived after the Blazhko modulations have been removed (see Section \ref{sec4} for more details).}
  %\tablecomments{}
\end{deluxetable*}

\begin{figure}
  \epsscale{1.1}
  \plotone{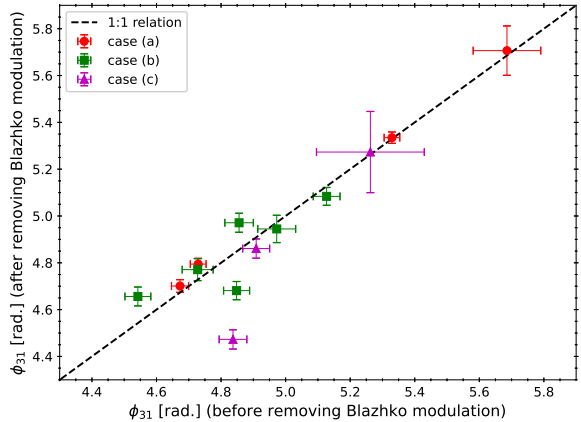}
  \caption{Comparison of the derived $V$-band $\phi^s_{31}$ Fourier parameters before and after removing the Blazhko modulations from the light curves. The filled color symbols corresponding to the three cases as presented in Figure \ref{fig_blazhko}. The dashed line represent the $1:1$ relation.}
  \label{fig_ph31}
\end{figure}

In general, light-curve for a periodic variable star can be fitted with an $n$-order Fourier expansion in the following form \cite[for example, see][]{petersen1986,deb2009}:

\begin{eqnarray}
  m(\Phi) & = & m_0 + \sum^n_{i=1} \left[ a_i \cos (2 \pi i  \Phi) + b_i \sin (2 \pi i  \Phi)\right],
\end{eqnarray}

\noindent where $\Phi=t/P-INT(t/P)$\footnote{$INT(x)$ is a function that returns the integer of $x$. Normally $t$ will be subtracted with a reference epoch $t_0$ (for example, at maximum light). For simplicity we assume $t_0=0$.} is the pulsational phases (between 0 and 1) after folding a time-series $t$ with $P$. With trigonometric identities, equation (8) can be re-written either as a sine series or a cosine series, i.e. in the form of $m(\Phi) = m_0 + \sum^n_{i=1} A_i \sin (2 \pi i \Phi + \phi^s_i)$ or $m(\Phi) = m_0 + \sum^n_{i=1} A_i \cos (2 \pi i \Phi + \phi^c_i)$. Following \citet{simon1981}, the Fourier parameter $\phi_{31}$ is either defined as $\phi^s_{31}=\phi^s_3 -3\phi^s_1$ or $\phi^c_{31}=\phi^c_3 -3\phi^c_1$, with a conversion of $\phi^c_{31}=\phi^s_{31}-\pi$ \citep{ngeow2016}. To be consistent with \citet{mullen2021}, we adopted the sine series and $n=5$ for obtaining the Fourier parameter $\phi^s_{31}$. Since the error on $\phi_{31}$ is the same for either the sine series or the cosine series, we calculated the error on $\phi^s_{31}$ using the prescription given in \citet{petersen1986} and \citet{petersen1994}.%, and propagated to the error of \feh$_V$.

For the 13 Blazhko RR Lyrae, we have also removed the modulated components following a similar procedures as described in \citet[][in their Section 5.2; also see the reference therein]{ngeow2016}, where the modulated (or Blazhko) periods, $P_{BL}$, were adopted from \citet{nemec2013}. The difference between this work and \citet{ngeow2016} is we fixed $n=5$, and only varying the $(r,\ q)$ Fourier orders when fitting the modulated components. The Fourier order $r$ and $q$ are similar to equation (8) but for the modulated frequency $f_m = 1/P_{BL}$, and the combined frequencies $kf_0 \pm f_m$ (where $f_0=1/P$ is the pulsation frequency, and $k$ is an integer runs from 0 to $q$), respectively. The same $(r,\ q)$ Fourier orders were adopted to fit the $grV$ and $VT$ light-curves for a given Blazhko RR Lyrae. We found that there are four Blazhko RR Lyrae (KIC 5559631, 10789273, 11125706, and 12155928) did not show Blazhko modulation on their light-curves. In contrast, improvements can be seen on the light-curves for six Blazhko RR Lyrae (KIC 6183128, 7505345, 7671081, 9001926, 9578833, and 9697825) after removing the Blazhko modulation. The remaining three Blazhko RR Lyrae (KIC 3864443, 4484128, and 7198959) we cannot effectively remove the Blazhko modulation for a variety of $(r,\ q)$ combinations. Examples of these three cases are presented in Figure \ref{fig_blazhko}. After removing the Blazhko modulation of the 13 Blazhko RR Lyrae (including the three RR Lyrae at which their Blazhko modulations cannot be effectively removed), we re-determined their Fourier parameter $\phi^s_{31}$ from the non-modulated light-curves. As an example, Figure \ref{fig_ph31} compares the $V$-band $\phi^s_{31}$ Fourier parameters before and after removing the Blazhko modulations. All of the derived $\phi^s_{31}$ for the 30 RR Lyrae are summarized in Table \ref{tab_fourier}. 

\section{Testing the Relation} \label{sec5}

\begin{figure*}
  \epsscale{1.1}
  \plottwo{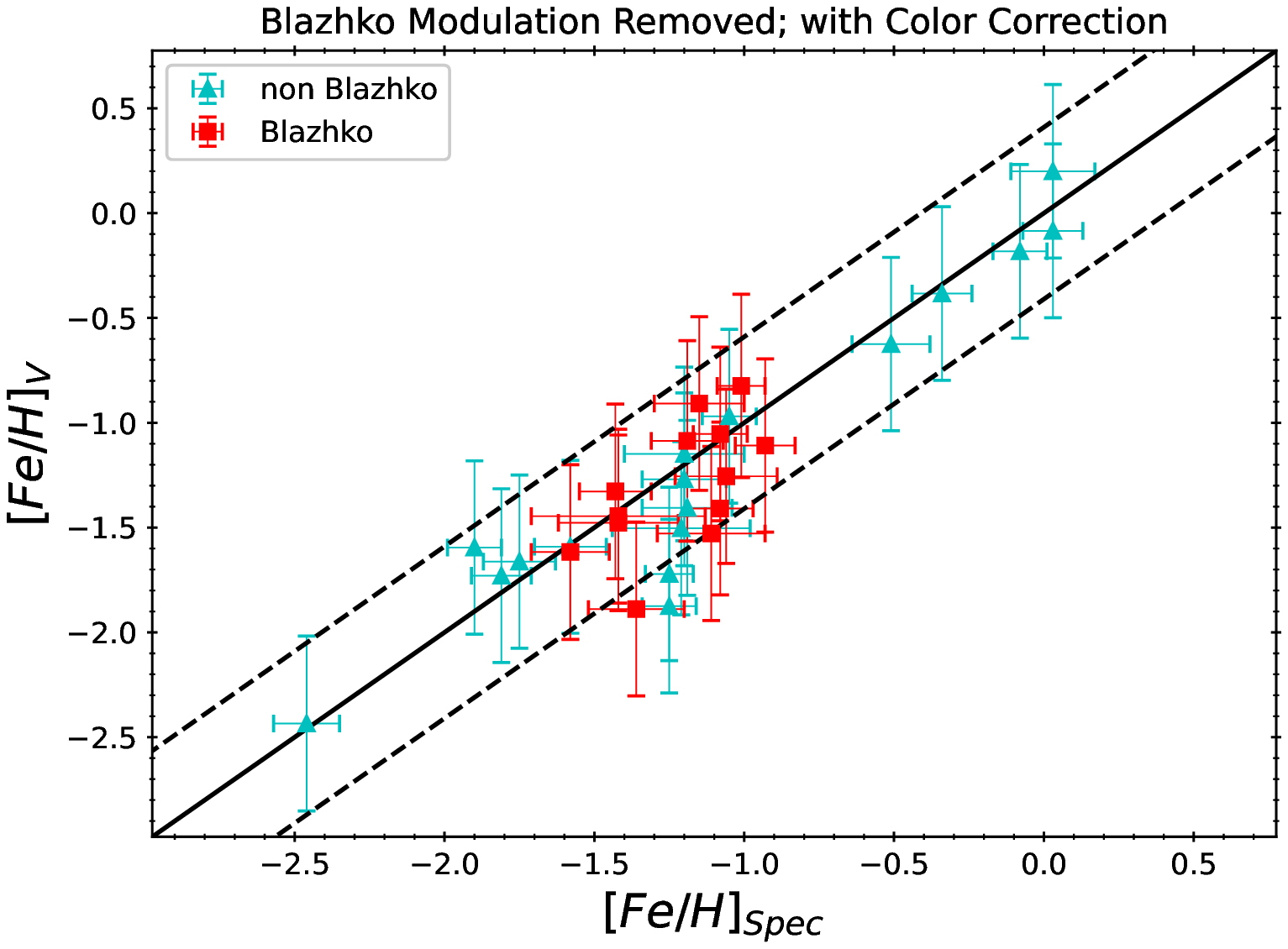}{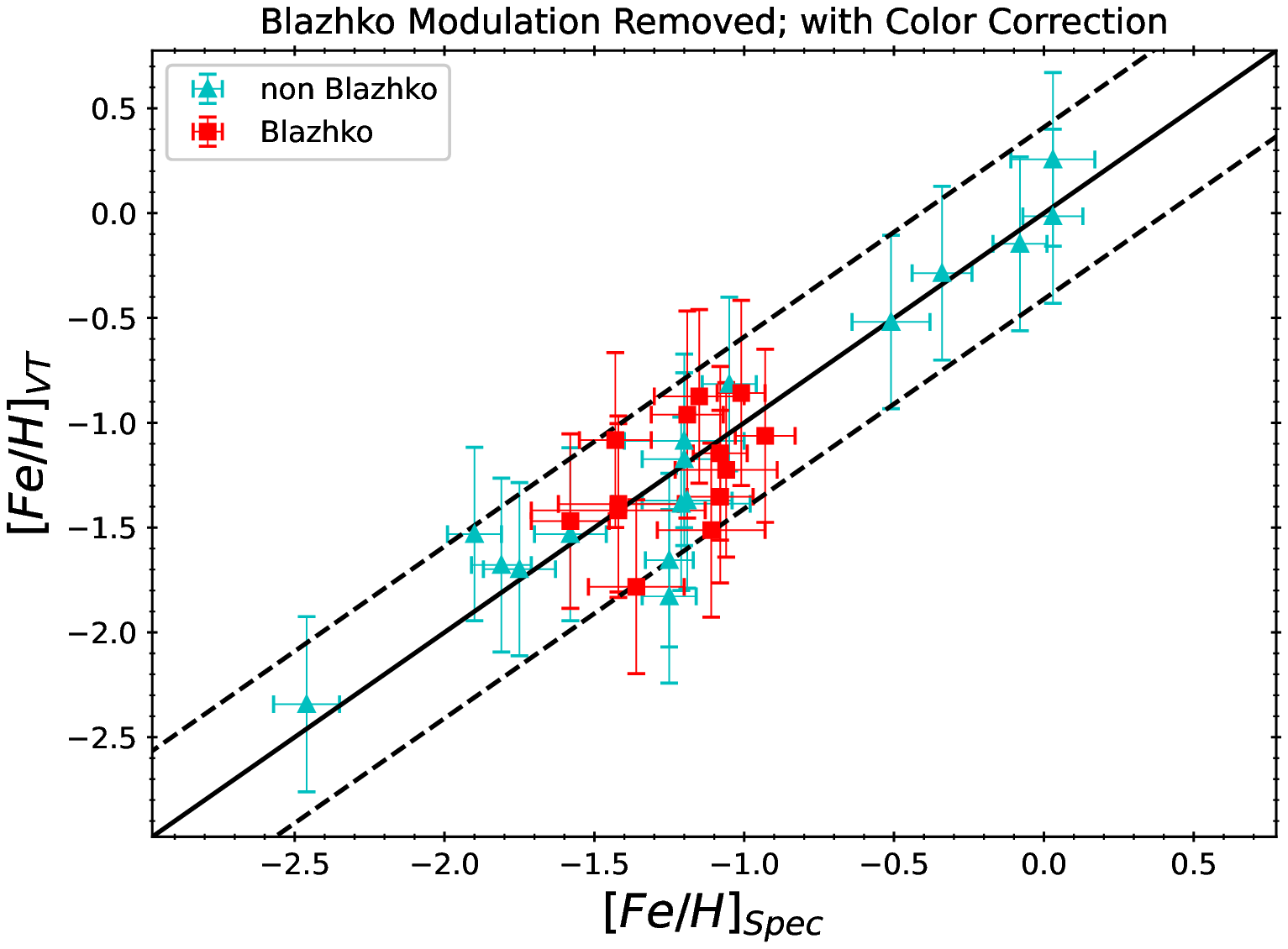}
  \plottwo{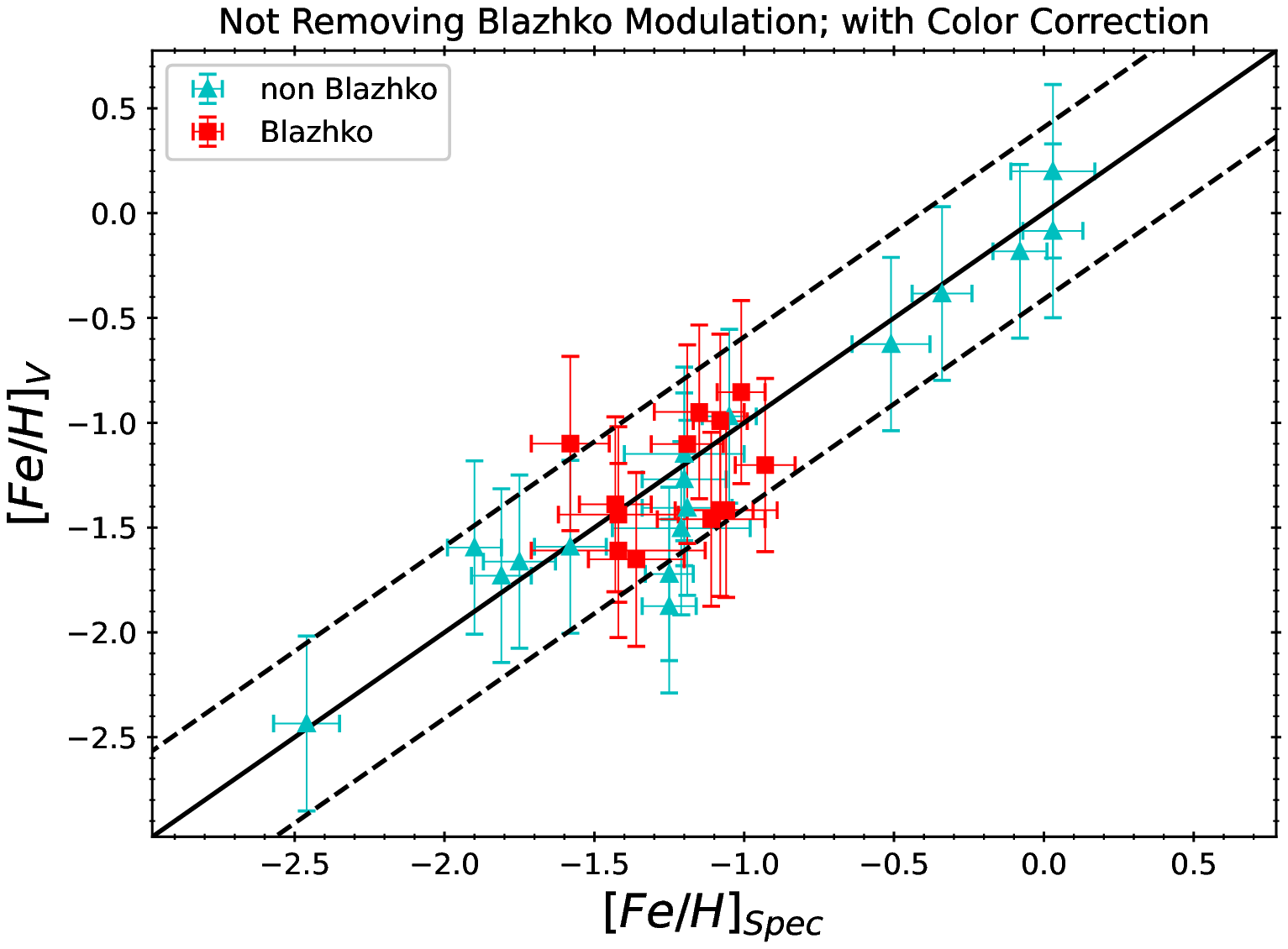}{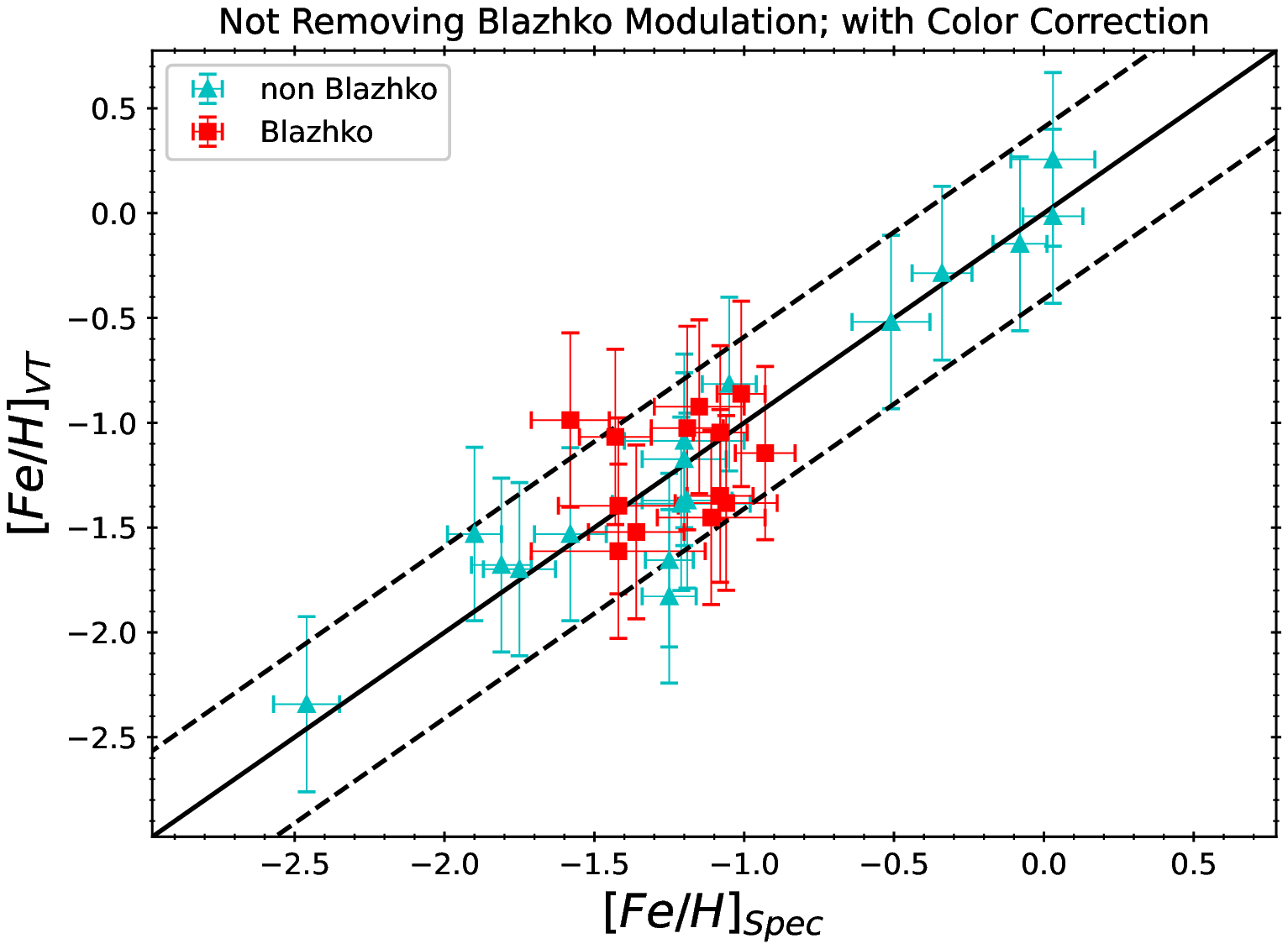}
  \plottwo{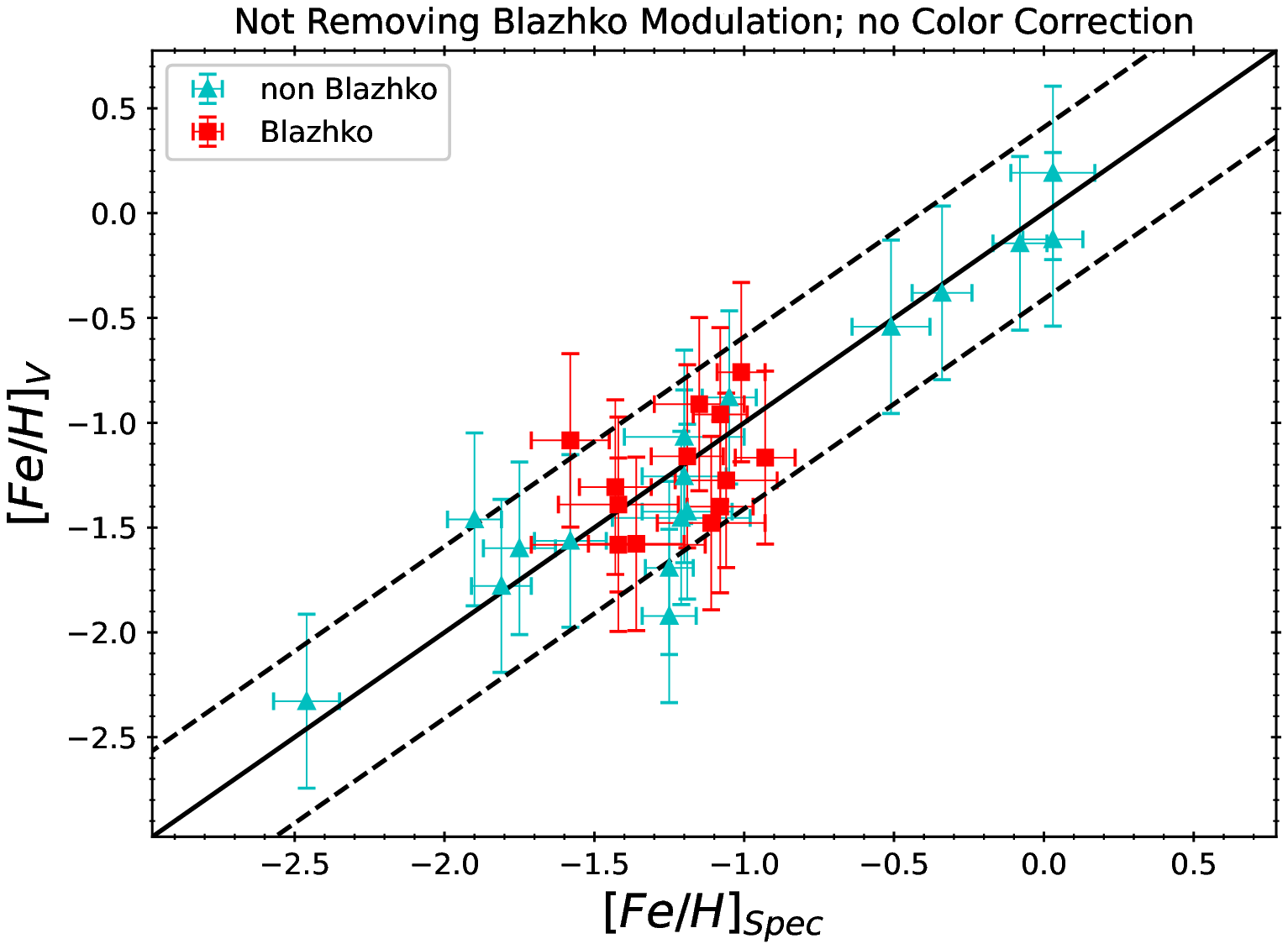}{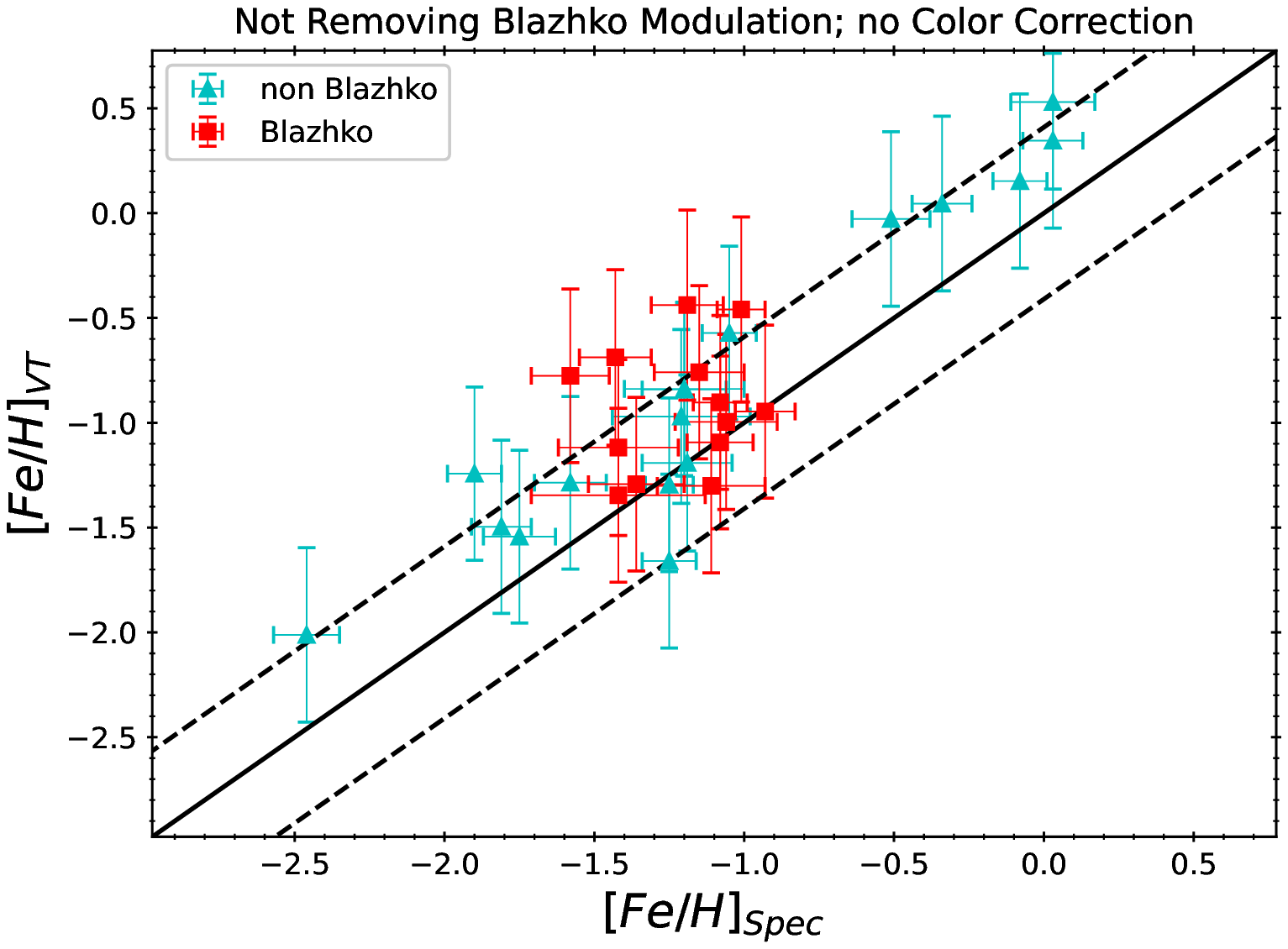}
  \caption{Comparisons of the photometric \feh, either in the $V$- (left panels) or $VT$-band (right panels), derived from equation (7) to the \feh$_{\mathrm{spec}}$. Errors on photometric \feh~include the RMS error and propagated errors on $\phi_{31}$ from equation (7), and we assume errors on $P$ are negligible. The top, middle, and bottom panels are the comparisons for three different cases as discussed in the text (see Section \ref{sec5}). The solid lines represent the $1:1$ relation and they are {\it not} the fits to the data. The dashed lines show the expected $\pm0.41$~dex RMS on the photometric \feh.}
  \label{fig_case}
\end{figure*}

In this Section, we test the performance of derived photometric \feh~from the $V$-band and the $VT$-band in several cases, assuming the spectroscopic \feh, \feh$_{\mathrm{spec}}$ adopted from \citet{nemec2013}, is the ``ground truth''. Since the \feh$_{\mathrm{spec}}$ from \citet{nemec2013} and the \feh$_V$ based on equation (7) are in \citet{carretta2009} and \citet{crestani2021} scale, respectively, we added an offset of $+0.08$~dex \citep[as determined in][]{mullen2021} to the \feh$_{\mathrm{spec}}$ when comparing these two \feh~values.

We first compare the \feh$_V$ calculated from equation (7) and \feh$_{\mathrm{spec}}$ using the calibrated or transformed light-curves, and after removing the Blazhko modulation for the Blazhko RR Lyrae, in the upper panels of Figure \ref{fig_case}. Good agreements between \feh$_V$ and \feh$_{\mathrm{spec}}$ can be seen from the upper left panel of Figure \ref{fig_case}, with $\langle$\feh$_V-$\feh$_{\mathrm{spec}}\rangle=-0.08$~dex. Similarly, photometric \feh~from the transformed $VT$-band light-curves agree well with the \feh$_{\mathrm{spec}}$, with $\langle$\feh$_{VT}-$\feh$_{\mathrm{spec}}\rangle=-0.01$~dex. In both cases, the standard deviations ($\sigma$) on the averaged values are $\sigma=0.23$~dex and $\sigma=0.24$~dex, respectively, well within the RMS of 0.41~dex given in equation (7).

Since the Blazhko periods might not be reliable determined, or might not be determined at all, for (possible) Blazhko RR Lyrae found in the synoptic time-series imaging surveys, we tested the $V$-band photometric \feh~if the Blazhko modulations were not removed on these Blazhko RR Lyrae. The middle panels of Figure \ref{fig_case} are similar to the upper panels, except that the Blazhko modulations were not removed. For the 13 Blazhko RR Lyrae with Blazhko modulations removed, we found $\langle$\feh$_V-$\feh$_{\mathrm{spec}}\rangle=-0.09$~dex ($\sigma = 0.23$~dex) and $\langle$\feh$_{VT}-$\feh$_{\mathrm{spec}}\rangle=-0.02$~dex ($\sigma = 0.25$~dex). In comparison, the averaged values changed to $-0.06$~dex ($\sigma = 0.26$~dex) and $0.00$~dex ($\sigma = 0.29$~dex) in the $V$- and $VT$-band, respectively, if the Blazhko modulations were not removed. Nevertheless, these averaged differences (and their $\sigma$) are well within the RMS of equation (7). Therefore, it is possible to include the Blazhko RR Lyrae in estimating the $V$-band photometric \feh~without removing their Blazhko modulations. A similar conclusion was also found in \citet{ngeow2016}.

Finally, we consider an extreme case such that the Blazhko modulations were not removed for the Blazhko RR Lyrae, at the same time the color corrections were ignored (see bottom panels of Figure \ref{fig_case}). That is, $(B-V)=0.0$~mag and $(g^{PS1}-r^{PS1})=0.0$~mag in equation (3) and (5) when calibrating the $V-$ and $VT$-band light-curves, hence $V=V^{\mathrm{instr}}+ZP_V$ and $VT=r^{\mathrm{instr}}+ZP_r + 0.006$. As a result, the $VT$-band light-curves are equivalent to $r$-band light-curves. In this case the averaged difference for the $V$-band photometric \feh~is $-0.03$~dex ($\sigma = 0.25$~dex), which is still reasonable to apply equation (7) to estimate the photometric \feh. In contrast, the $VT$-band photometric \feh~display a large offset from the $1:1$ relation in the bottom right panel of Figure \ref{fig_case}, with a much larger averaged difference of $0.28$~dex ($\sigma = 0.28$~dex). Even though this value is still within the 0.41~dex RMS of equation (7), it is large enough to induce a bias in the derived photometric \feh. When the color corrections were ignored or set to zero, $\phi_{31}$ values derived from the $VT$-band light curves are equivalent to those from the $r$-band light-curves, hence equation (7) should not be used.

Would it be possible to use a mean color when transforming the $gr$-band light curves to the $VT$-band light curves? We tested this scenario by using the mean colors for each RR Lyrae, and repeated the same procedures. In this case $\langle$\feh$_{VT}-$\feh$_{\mathrm{spec}}\rangle=0.28$~dex with $\sigma=0.29$~dex. This scenario is similar to the previous case for setting the color to zero, as the constant color-term can be ``absorbed'' to the $ZP_m$, hence the $VT$-band light-curves would be similar to the $r$-band light curves. To remedy this, we provided a template color-curve in Appendix \ref{appxb}, such that colors can be estimated at various pulsational phases using the template color-curve and an estimation of the mean color, and not assuming a constant or zero color. There are various approaches to estimate the mean color for an ab-type RR Lyrae, such as using the observed light-curves, using prior information (e.g. from other surveys or observations), using a period-color relation \cite[e.g., in][]{ngeow2022}, etc. It is up to the researchers to decide which approach to use, depending on their situations, needs and goals.

\section{Deriving the Relations} \label{sec6}

\begin{figure*}
  \gridline{
    \fig{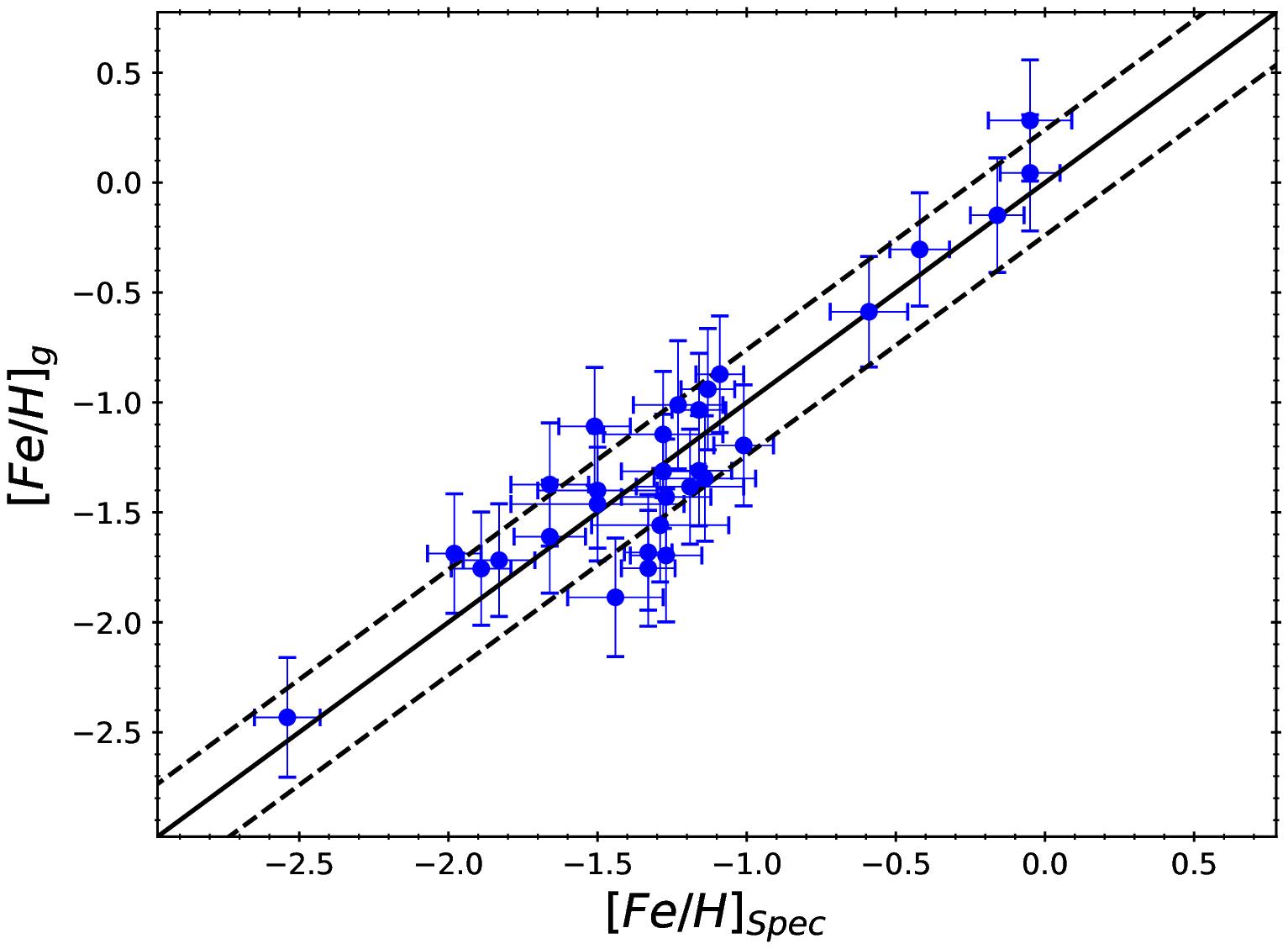}{0.32\textwidth}{(a) \feh~using equation (9)}
    \fig{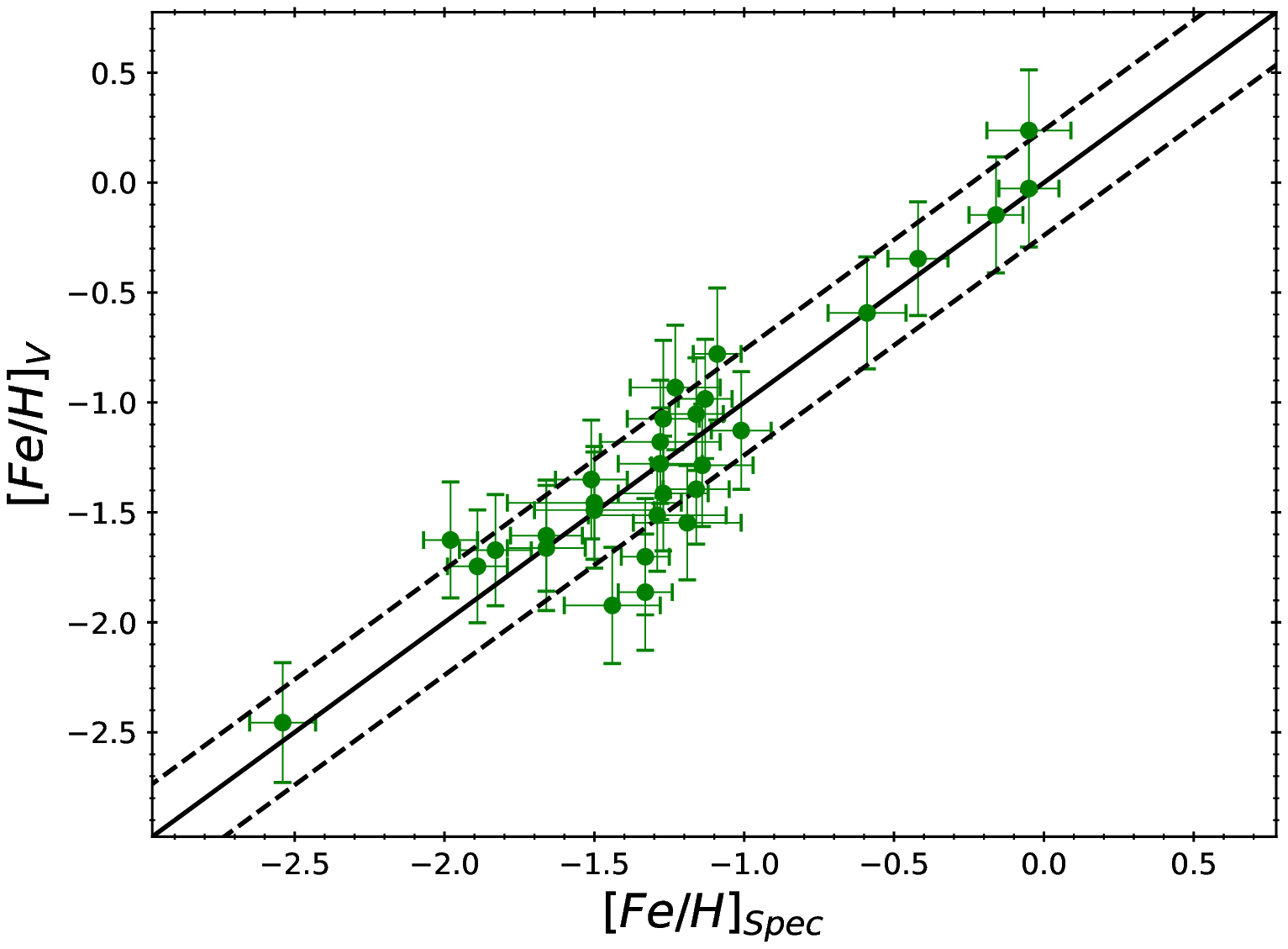}{0.32\textwidth}{(b) \feh~using equation (10)}
    \fig{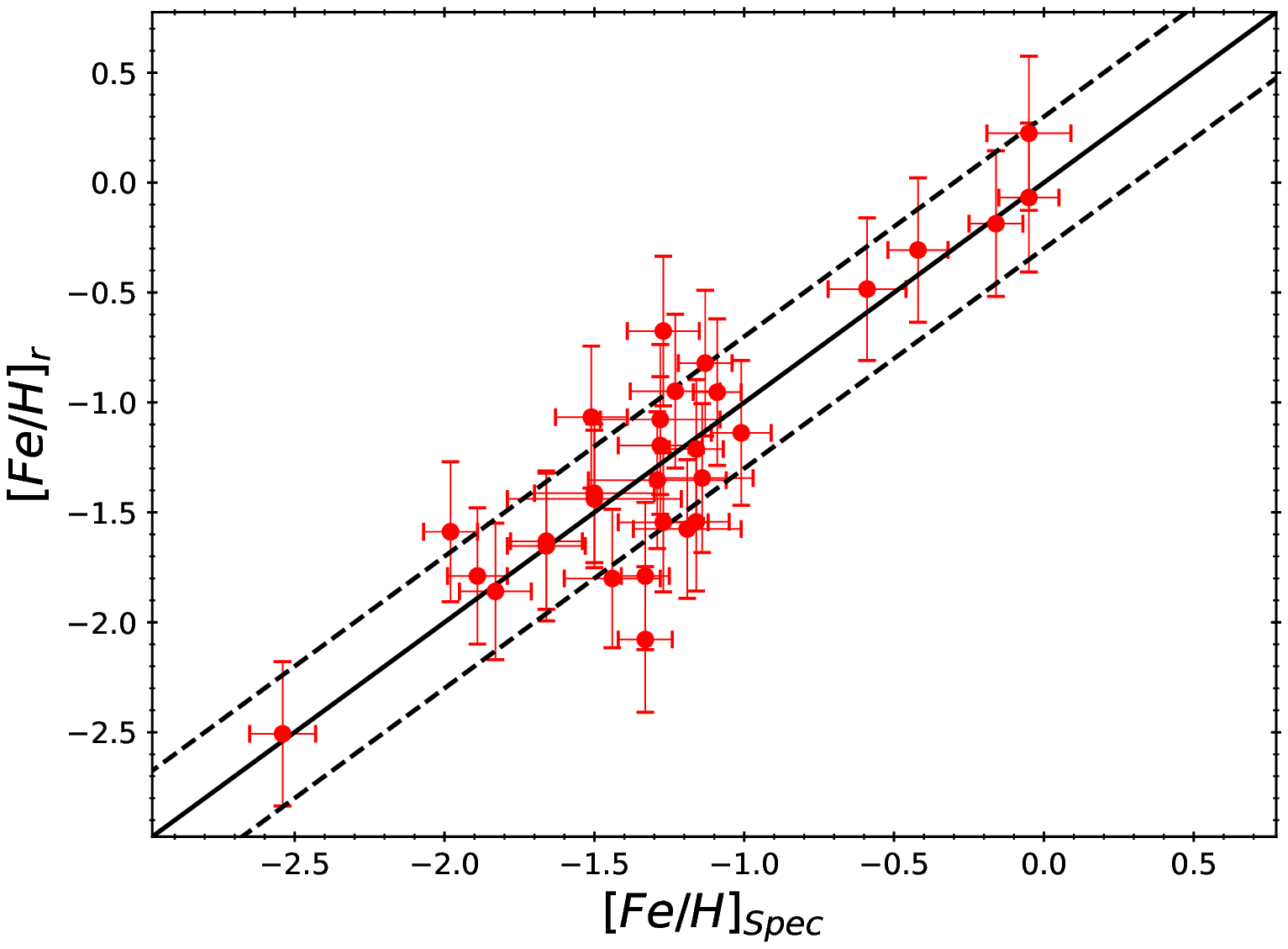}{0.32\textwidth}{(c) \feh~using equation (11) }
  }
  \caption{Comparisons of the predicted $gVr$-band photometric \feh, using equation (9), (10), and (11), to the \feh$_{\mathrm{spec}}$ for all of the 30 ab-type RR Lyrae listed in Table \ref{tab_fourier}. The solid lines are the $1:1$ relations, and the dashed lines represent the expected RMS given in equation (9), (10), and (11).}
  \label{fig_grv}
\end{figure*}

The Fourier parameter $\phi_{31}$ presented in Table \ref{tab_fourier} for all of the 30 ab-type RR Lyrae can be used to derive independent \feh-$\phi_{31}$-$P$ relation in the $gVr$ bands. Following \citet{mullen2021}, we adopted a  \feh-$\phi_{31}$-$P$ relation in the form of \feh$=a+b(P-P_0)+c(\phi_{31}-\phi_{31}^0)$, where $P_0$ and $\phi_{31}^0$ are the mean period and $\phi_{31}$ in the sample. The zero-point ($a$), the period coefficient ($b$), and the $\phi_{31}$ coefficient ($c$) were fitted using the orthogonal distance regression (ODR) implemented in the {\tt SciPy} package (i.e. {\tt scipy.odr}), at which errors on both of the \feh~and $\phi_{31}$ were included in the fittings. The mean period for our sample is $P_0=0.55$~days, and the mean $\phi_{31}^0$ values in the $gVr$-band are 4.97, 5.06, and 5.27~rad, respectively. To be consistent with \citet{mullen2021}, we adopted $P_0=0.58$~days and $\phi_{31}^0=5.25$~rad, and derived the following relations:

\begin{eqnarray}
  \mathrm{[Fe/H]_g}  & = & -1.01[\pm0.06] - 7.43[\pm0.80] (P-0.58) + 1.69 \nonumber \\
   & & [\pm0.16] (\phi_{31}^s - 5.25), \ \ RMS = 0.24\ \mathrm{dex}, \\
  \mathrm{[Fe/H]_V}  & = & -1.21[\pm0.05] - 7.67[\pm0.78] (P-0.58) + 1.50 \nonumber \\
  & & [\pm0.14] (\phi_{31}^s - 5.25), \ \ RMS = 0.24\ \mathrm{dex}, \\
  \mathrm{[Fe/H]_r}  & = & -1.54[\pm0.07] - 8.28[\pm1.04] (P-0.58) + 1.30 \nonumber \\
  & & [\pm0.15] (\phi_{31}^s - 5.25), \ \ RMS = 0.30\ \mathrm{dex}. 
\end{eqnarray}

\noindent If we adopted $P_0$ and $\phi_{31}^0$ from our sample, then the zero-points ($a$) were changed to $-1.26\pm0.04$, $-1.26\pm0.04$, and $-1.27\pm0.05$ in the $gVr$-band, respectively. We emphasize that equation (9) to (11) are only applicable to ab-type RR Lyrae.

{Similar to \citet{nemec2013} and \citet{ngeow2016}, we compare the predicted photometric \feh~derived from equation (9), (10), and (11) to the \feh$_{\mathrm{spec}}$ in Figure \ref{fig_grv}. The Fourier parameters $\phi_{31}$ were adopted from Table \ref{tab_fourier}, including the Blazhko RR Lyrae with Blazhko modulations being removed, when calculating the predicted photometric \feh. All of the three equations give $\langle$\feh$_{g,V,r}-$\feh$_{\mathrm{spec}}\rangle=0.0$~dex, with $\sigma_g=0.24$~dex, $\sigma_V=0.23$~dex, and $\sigma_r=0.29$~dex, respectively. These standard deviations are the same or similar to the RMS given in equation (9) to (11), and no systematic offsets were seen from Figure \ref{fig_grv}.

Recall that the \feh$_{g,V,r}$ in equation (9) to (11) are in the \citet{carretta2009} scale. If the \feh~were converted to the \citet{crestani2021} scale, same as in \citet{mullen2021}, then the $gVr$-band zero-points ($a$) of the relation became $-0.93\pm0.06$, $-1.13\pm0.05$, and $-1.46\pm0.07$, respectively.

\section{Discussions and Conclusions} \label{sec7}

In this work, we tested the performance of \feh$_V$ from \citet{mullen2021} using a set of homogeneous ab-type RR Lyrae located in the {\it Kepler} field. The most important conclusion from our work is good agreement was found between \feh$_V$ and \feh$_{\mathrm{spec}}$. This conclusion also holds if the $V$-band light-curves are photometrically transformed from the $gr$-band light-curves, as long as the color-terms are properly taken into account when calibrating the photometry. On the other hand, the transformed $V$-band (i.e. the $VT$-band) light-curves should not be used to derive \feh$_V$ if the color-terms are ignored or set to a constant. Modern synoptic time-series surveys often will not acquire near simultaneous $g$- and $r$-band photometry\footnote{Recall that RR Lyrae are short period pulsating stars, even the $g$- and $r$-band photometry were taken in the same night but separated in few hours cannot be treated as ``near simultaneous''.}, hence estimating the $(g-r)$ colors at certain pulsational phases can be a non-trivial task. In Appendix \ref{appxb}, we provide a template color-curve such that the $(g-r)$ colors can be estimated at various pulsational phases.

Our sample also contains $\sim40\%$ Blazhko RR Lyrae, providing an opportunity to test the inclusion of Blazhko RR Lyrae in deriving \feh$_V$. Our test results suggested Blazhko RR Lyrae can be used even the Blazhko modulations are not removed. This implies RR Lyrae found in the synoptic time-series surveys \citep[for example the LSST in coming years, see][]{hernischek2022} can be used to determine the photometric \feh~without first identifying if they are Blazhko RR Lyrae or not.

Finally, we derive independent \feh-$\phi_{31}$-$P$ relations, for the ab-type RR Lyrae, in the $gVr$ bands based on our data. Good agreement between the coefficients can be seen in the $V$-band \feh-$\phi_{31}$-$P$ relation derived in \citet{mullen2021} and in our work, suggesting the $V$-band \feh-$\phi_{31}$-$P$ relation is robust, no matter it is derived from an ``all-sky'' sample as done in \citet[][with $\sim 10^3$ ab-type RR Lyrae]{mullen2021} or from a ``local'' sample of 30 ab-type RR Lyrae located in the {\it Kepler} field. Furthermore, the $V$-band light-curves are totally independent in these works. Together with the infra-red \feh-$\phi_{31}$-$P$ relation presented in \citet{mullen2021}, i.e. \feh$_{WISE}=-1.47-8.33(P-0.58)+0.92(\phi_{31}-1.90)$}, we found that both period coefficient ($b$) and $\phi_{31}$ coefficient ($c$) in the \feh-$\phi_{31}$-$P$ relations monotonically decrease when the wavelength is increasing.

\acknowledgments

We thank the useful discussions and comments from an anonymous referee to improve the manuscript. We sincerely thank the observing staff at the Lulin Observatory, C.-S. Lin, H.-Y. Hsiao, and W.-J. Hou, to carry out the queue observations for this work. We are thankful for funding from the Ministry of Science and Technology (Taiwan) under the contract 109-2112-M-008-014-MY3. This research made use of {\tt Astropy},\footnote{\url{http://www.astropy.org}} a community-developed core Python package for Astronomy \citep{astropy2013, astropy2018}. This research has made use of the VizieR catalogue access tool, CDS, Strasbourg, France. The original description of the VizieR service was published in \citet{ochsenbein2000}.

The Pan-STARRS1 Surveys (PS1) and the PS1 public science archive have been made possible through contributions by the Institute for Astronomy, the University of Hawaii, the Pan-STARRS Project Office, the Max-Planck Society and its participating institutes, the Max Planck Institute for Astronomy, Heidelberg and the Max Planck Institute for Extraterrestrial Physics, Garching, The Johns Hopkins University, Durham University, the University of Edinburgh, the Queen's University Belfast, the Harvard-Smithsonian Center for Astrophysics, the Las Cumbres Observatory Global Telescope Network Incorporated, the National Central University of Taiwan, the Space Telescope Science Institute, the National Aeronautics and Space Administration under Grant No. NNX08AR22G issued through the Planetary Science Division of the NASA Science Mission Directorate, the National Science Foundation Grant No. AST-1238877, the University of Maryland, Eotvos Lorand University (ELTE), the Los Alamos National Laboratory, and the Gordon and Betty Moore Foundation.

\software{{\tt astropy} \citep{astropy2013,astropy2018}, {\tt astrometry.net} \citep{lang2010}, {\tt IRAF} \citep{tody1986,tody1993}, {\tt SExtractor} \citep{bertin1996}, {\tt Matplotlib} \citep{hunter2007},  {\tt NumPy} \citep{harris2020}, {\tt SciPy} \citep{virtanen2020}.}

%\newpage

\appendix

\section{Selection of Reference Stars} \label{appxa}

\begin{figure*}
  \epsscale{1.15}
  \plotone{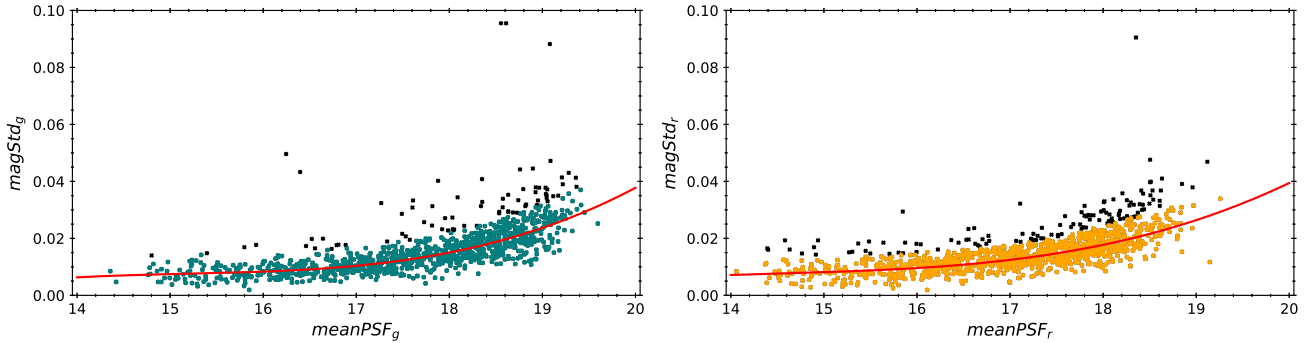}
  \caption{An example of the plots of $magStd$ vs. $meanPSF$ in the $g$- and $r$-band, shown in the left and right panel, respectively, for KIC12155928. The black points are the rejected data points after the initial fitting with a $3^{\mathrm{rd}}$-degree polynomial function. The polynomial function was applied again to fit the remaining data points (in various colors) in the second step of the two-steps fitting procedure, and displayed as the red curves. Only the data points below the red curves were selected as reference stars. All data points were extracted from the Pan-STARRS1 DR1 photometric catalogs.}\label{fig_magstd}
\end{figure*}

We summarize a number of selection criteria to select non-varying stellar sources from the catalogs merged with Pan-STARRS1 DR1 photometric catalog and the $UBV$ photometric catalog given in \citet{everett2012}. These non-varying stellar sources will be used to construct a reference stars catalog for each of the targeted RR Lyrae. The selection criteria are listed below, some of them were inspired from, or similar to those adopted in, the ZTF Science Data System Explanatory Supplement.\footnote{See \url{https://irsa.ipac.caltech.edu/data/ZTF/docs/ztf\_explanatory\_supplement.pdf}}

\begin{itemize}
\item Excluding the targeted RR Lyrae.
\item Separation of the matched sources when cross-matching the Pan-STARRS1 DR1 and the $UBV$ catalogs is smaller than $1\arcsec$.
\item Sources without measurements in any of the $BVgri$ bands in the merged catalogs were excluded.
\item Sources with number of observations in each of the $gri$-band is greater than 5.
\item Sources with mean PSF magnitudes ($meanPSF$) in $gri$-band that are in between $14$~mag and $20$~mag.
\item Sources with difference between the $i$-band mean PSF magnitudes and mean Kron magnitudes ($meanKron$), $meanPSF_i - meanKron_i$, that are within the range of $-1.5$~mag and $-0.04$~mag.
\item Sources with color, $meanPSF_g - meanPSF_i$, that are within the range of $-0.5$~mag and $3.0$~mag.
\item Sources with photometric errors in all of the $BVgr$ bands that are smaller than $0.1$~mag.
\item Sources with standard deviations ($magStd$) in the $gr$-band $meanPSF$ that are smaller than $0.1$~mag.
\end{itemize}

\noindent For the remaining sources after applying the above selection criteria, a $3^{\mathrm{rd}}$-degree polynomial function was fitted to the $gr$-band $meanPSF$ and $magStd$ in a two-steps process. In the first step, we removed data points that were larger than 1 standard deviation from the best-fit polynomial function. The rest of the data points were re-fit in the second step of the process (see Figure \ref{fig_magstd}). We only retained sources that have $magStd$ smaller than the best-fit polynomial function in both $gr$-band as reference stars for performing the photometric calibration (see Section \ref{sec3}). The number of the selected reference stars varies from $\sim 160$ to $\sim 1120$ for the 30 targeted RR Lyrae.

\section{Color-Curve Template} \label{appxb}

\begin{figure*}
  \epsscale{1.1}
  \plottwo{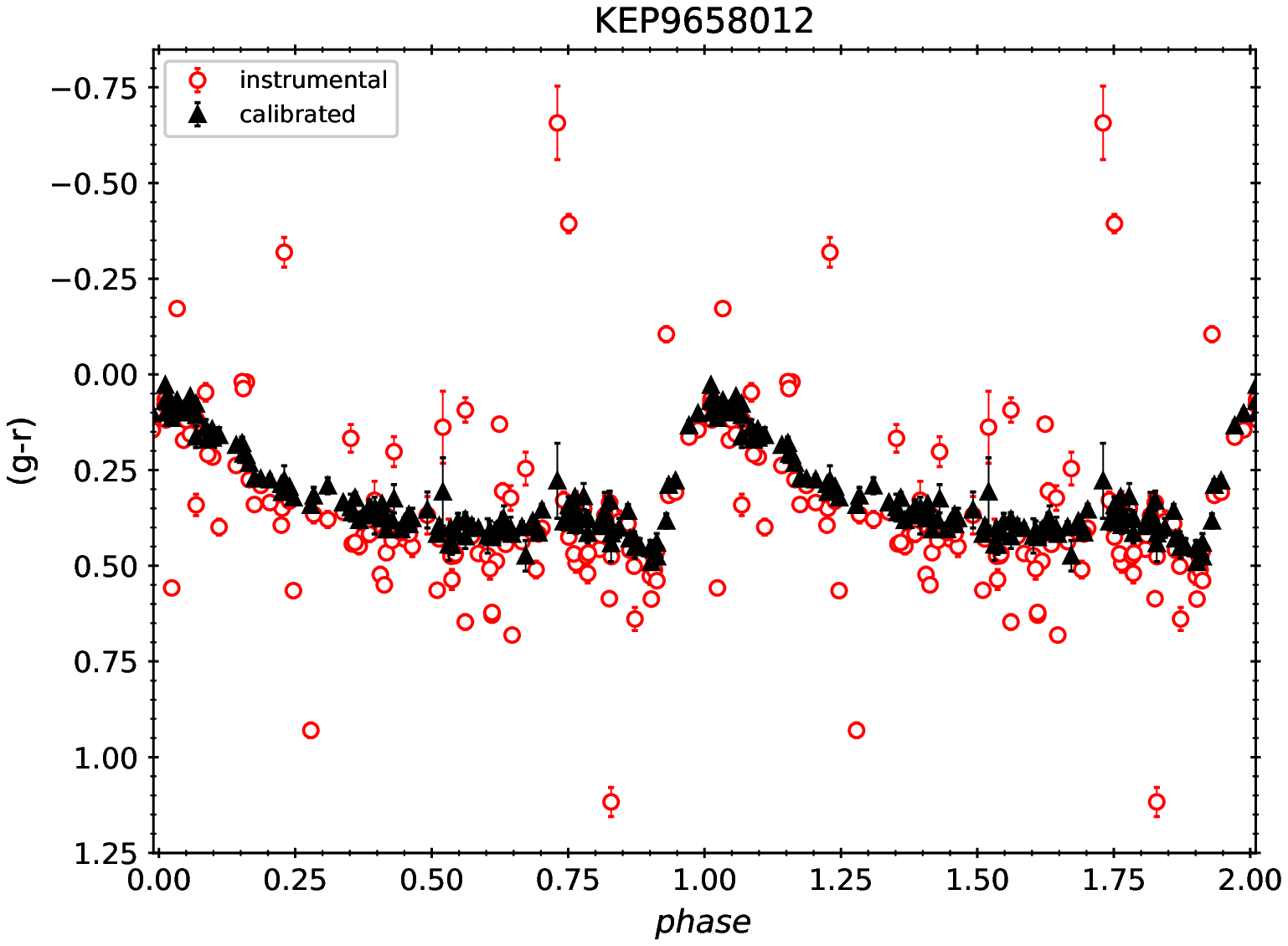}{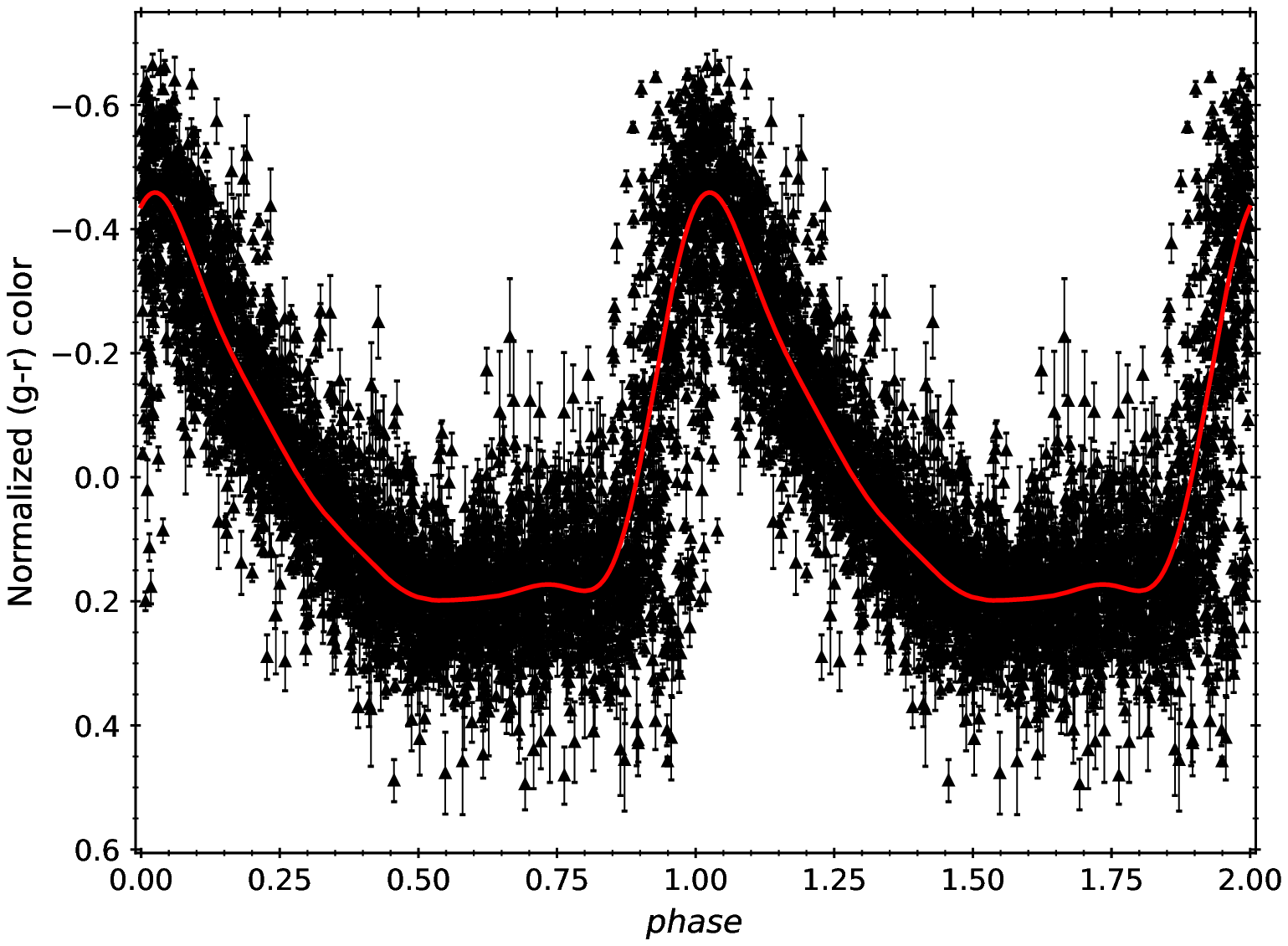}
  \caption{{\bf Left panel:} Comparison of the instrumental $(g^{\mathrm{instr}} - r^{\mathrm{instr}})$ colors (in red open circles) and the calibrated $(g^{PS1}-r^{PS1})$ colors (in black filled triangles), using equation (6), for the same RR Lyrae as presented in Figure \ref{fig_lc}. {\bf Right panel:} Composite of the normalized $(g-r)$ color-curve based on 27 RR Lyrae (excluding KIC 3864443, 4484128, and 7198959; due to larger scatters seen on their color-curves), the red curve represents the best-fit template color-curve (see text for details).}
  \label{fig_color}
\end{figure*}

The near simultaneous $gr$-band observations of our targeted RR Lyrae allow the construction of color-curves via equation (6), an example is presented in the left panel of Figure \ref{fig_color}. These color-curves can be used to construct a template color-curve in the $(g-r)$ color. We first removed data points with errors larger than $0.1$~mag (they tend to be outliers on the color-curves), and then normalized the color-curves by subtracting the means and scaled with their amplitudes. Finally, we re-phased the color-curves using the reference epoch $t_0$ adopted from \citet[][their Table 1]{nemec2013}. The composite color-curve is presented in the right panel of Figure \ref{fig_color}, and fitted with a low-order Fourier sine series. The best-fit template color-curve, shown as red curve in the right panel of Figure \ref{fig_color}, is:

\begin{eqnarray}
  c(\Phi) & = & 0.280\sin (2\pi \Phi + 4.205) + 0.111\sin (4\pi \Phi + 4.088) + 0.062\sin (6\pi \Phi + 4.430), \nonumber \\
   & & + 0.027 \sin (8\pi \Phi + 4.808) + 0.011 \sin (10\pi \Phi + 5.138),
\end{eqnarray}

\noindent with an RMS of 0.131, where $c=(g-r)$.

% ===============================================
%               REFERENCE
% ===============================================


\begin{thebibliography}{} % CHECK -- removed not cited

\bibitem[Aihara et al.(2018)]{aihara2018} Aihara, H., Arimoto, N., Armstrong, R., et al.\ 2018, \pasj, 70, S4

\bibitem[Alcock et al.(2000)]{alcock2000} Alcock, C., Allsman, R.~A., Alves, D.~R., et al.\ 2000, \aj, 119, 2194
  
\bibitem[Astropy Collaboration et al.(2013)]{astropy2013} Astropy Collaboration, Robitaille, T.~P., Tollerud, E.~J., et al.\ 2013, \aap, 558, A33

\bibitem[Astropy Collaboration et al.(2018)]{astropy2018} Astropy Collaboration, Price-Whelan, A.~M., Sip{\H{o}}cz, B.~M., et al.\ 2018, \aj, 156, 123

\bibitem[Beaton et al.(2018)]{beaton2018} Beaton, R.~L., Bono, G., Braga, V.~F., et al.\ 2018, \ssr, 214, 113

\bibitem[Bellm et al.(2019)]{bel19} Bellm, E.~C., Kulkarni, S.~R., Graham, M.~J., et al.\ 2019, \pasp, 131, 018002

\bibitem[Bertin \& Arnouts(1996)]{bertin1996} Bertin, E. \& Arnouts, S.\ 1996, \aaps, 117, 393
  
\bibitem[Bhardwaj(2020)]{bhardwaj2020} Bhardwaj, A.\ 2020, Journal of Astrophysics and Astronomy, 41, 23

\bibitem[Carretta et al.(2009)]{carretta2009} Carretta, E., Bragaglia, A., Gratton, R., et al.\ 2009, \aap, 508, 695
  
\bibitem[Chambers et al.(2016)]{chambers2016} Chambers, K.~C., Magnier, E.~A., Metcalfe, N., et al.\ 2016, arXiv:1612.05560

\bibitem[Crestani et al.(2021)]{crestani2021} Crestani, J., Fabrizio, M., Braga, V.~F., et al.\ 2021, \apj, 908, 20
  
\bibitem[Dark Energy Survey Collaboration et al.(2016)]{des2016} Dark Energy Survey Collaboration, Abbott, T., Abdalla, F.~B., et al.\ 2016, \mnras, 460, 1270

\bibitem[Deb \& Singh(2009)]{deb2009} Deb, S. \& Singh, H.~P.\ 2009, \aap, 507, 1729
  
\bibitem[D{\'e}k{\'a}ny et al.(2021)]{dekany2021} D{\'e}k{\'a}ny, I., Grebel, E.~K., \& Pojma{\'n}ski, G.\ 2021, \apj, 920, 33

\bibitem[Everett et al.(2012)]{everett2012} Everett, M.~E., Howell, S.~B., \& Kinemuchi, K.\ 2012, \pasp, 124, 316
  
\bibitem[Fabrizio et al.(2021)]{fabrizio2021} Fabrizio, M., Braga, V.~F., Crestani, J., et al.\ 2021, \apj, 919, 118

\bibitem[Flewelling et al.(2020)]{flewelling2020} Flewelling, H.~A., Magnier, E.~A., Chambers, K.~C., et al.\ 2020, \apjs, 251, 7
  
\bibitem[Graham et al.(2019)]{gra19} Graham, M.~J., Kulkarni, S.~R., Bellm, E.~C., et al.\ 2019, \pasp, 131, 078001
  
\bibitem[Harris et al.(2020)]{harris2020} Harris, C.~R., Millman, K.~J., van der Walt, S.~J., et al.\ 2020, \nat, 585, 357

\bibitem[Hernitschek \& Stassun(2022)]{hernischek2022} Hernitschek, N. \& Stassun, K.~G.\ 2022, \apjs, 258, 4
  
\bibitem[Hunter(2007)]{hunter2007} Hunter, J.~D.\ 2007, Computing in Science and Engineering, 9, 90
    
\bibitem[Iorio \& Belokurov(2021)]{iorio2021} Iorio, G. \& Belokurov, V.\ 2021, \mnras, 502, 5686

\bibitem[Ivezi{\'c} et al.(2019)]{ivezic2019} Ivezi{\'c}, {\v{Z}}., Kahn, S.~M., Tyson, J.~A., et al.\ 2019, \apj, 873, 111
  
\bibitem[Jurcsik \& Kovacs(1996)]{jurcsik1996} Jurcsik, J. \& Kovacs, G.\ 1996, \aap, 312, 111

\bibitem[Lang et al.(2010)]{lang2010} Lang, D., Hogg, D.~W., Mierle, K., et al.\ 2010, \aj, 139, 1782

\bibitem[Martinez-Vazquez et al.(2016)]{marvaz2016} Martinez-Vazquez, C.~E., Monelli, M., Bono, G., et al.\ 2016, Commmunications of the Konkoly Observatory Hungary, 105, 53

\bibitem[Masci et al.(2019)]{masci2019} Masci, F.~J., Laher, R.~R., Rusholme, B., et al.\ 2019, \pasp, 131, 018003
  
\bibitem[Morgan et al.(2007)]{morgan2007} Morgan, S.~M., Wahl, J.~N., \& Wieckhorst, R.~M.\ 2007, \mnras, 374, 1421

\bibitem[Mullen et al.(2021)]{mullen2021} Mullen, J.~P., Marengo, M., Mart{\'\i}nez-V{\'a}zquez, C.~E., et al.\ 2021, \apj, 912, 144
  
\bibitem[Nemec et al.(2011)]{nemec2011} Nemec, J.~M., Smolec, R., Benk{\H{o}}, J.~M., et al.\ 2011, \mnras, 417, 1022

\bibitem[Nemec et al.(2013)]{nemec2013} Nemec, J.~M., Cohen, J.~G., Ripepi, V., et al.\ 2013, \apj, 773, 181
  
\bibitem[Ngeow et al.(2016)]{ngeow2016} Ngeow, C.-C., Yu, P.-C., Bellm, E., et al.\ 2016, \apjs, 227, 30

\bibitem[Ngeow et al.(2022)]{ngeow2022} Ngeow, C.-C., Bhardwaj, A., Dekany, R., et al.\ 2022, \aj, 163, 239
  
\bibitem[Ochsenbein et al.(2000)]{ochsenbein2000} Ochsenbein, F., Bauer, P., \& Marcout, J.\ 2000, \aaps, 143, 23
  
\bibitem[Oluseyi et al.(2012)]{oluseyi2012} Oluseyi, H.~M., Becker, A.~C., Culliton, C., et al.\ 2012, \aj, 144, 9

\bibitem[Onken et al.(2019)]{onken2019} Onken, C.~A., Wolf, C., Bessell, M.~S., et al.\ 2019, \pasa, 36, e033

\bibitem[Petersen(1986)]{petersen1986} Petersen, J.~O.\ 1986, \aap, 170, 59

\bibitem[Petersen(1994)]{petersen1994} Petersen, J.~O.\ 1994, \aaps, 105, 145
  
\bibitem[Sandage(2004)]{sandage2004} Sandage, A.\ 2004, \aj, 128, 858
  
\bibitem[Sandage \& Tammann(2006)]{sandage2006} Sandage, A. \& Tammann, G.~A.\ 2006, \araa, 44, 93
  
\bibitem[Sesar et al.(2010)]{sesar2010} Sesar, B., Ivezi{\'c}, {\v{Z}}., Grammer, S.~H., et al.\ 2010, \apj, 708, 717

\bibitem[Simon \& Lee(1981)]{simon1981} Simon, N.~R. \& Lee, A.~S.\ 1981, \apj, 248, 291
  
\bibitem[Smolec(2005)]{smolec2005} Smolec, R.\ 2005, \actaa, 55, 59
  
\bibitem[Tody(1986)]{tody1986} Tody, D.\ 1986, \procspie, 627, 733

\bibitem[Tody(1993)]{tody1993} Tody, D.\ 1993, Astronomical Data Analysis Software and Systems II, 52, 173

\bibitem[Tonry et al.(2012)]{tonry2012} Tonry, J.~L., Stubbs, C.~W., Lykke, K.~R., et al.\ 2012, \apj, 750, 99
  
\bibitem[Virtanen et al.(2020)]{virtanen2020} Virtanen, P., Gommers, R., Oliphant, T.~E., et al.\ 2020, Nature Methods, 17, 261

\bibitem[Watkins et al.(2009)]{watkins2009} Watkins, L.~L., Evans, N.~W., Belokurov, V., et al.\ 2009, \mnras, 398, 1757
  
\bibitem[Wu et al.(2006)]{wu2006} Wu, C., Qiu, Y.~L., Deng, J.~S., et al.\ 2006, \aap, 453, 895
\end{thebibliography}
\end{document}